\documentclass{emulateapj}

\newcommand{\hmpc}{\,h^{-1}\,{\rm Mpc} }

\newcommand{\lta}{\lesssim}
\newcommand{\gta}{\gtrsim}

\newcommand{\SPH}{SPH}
\newcommand{\MMH}{MMH}
\newcommand{\COBE}{{\it COBE}--\textsc{DMR}}
\newcommand{\DMR}{DMR}
\newcommand{\Boomerang}{BOOMERANG}
\newcommand{\Maxima}{MAXIMA}
\newcommand{\DASI}{DASI}
\newcommand{\Toco}{\textsc{TOCO}}

\newcommand{\CBI}{\textsc{CBI}}

\begin{document}

\journalinfo{Accepted for publication in The Astrophysical Journal}
\submitted{}

\title{The Sunyaev-Zeldovich Effect in CMB-calibrated theories applied
to the Cosmic Background Imager anisotropy power at $\ell>2000$}

\shortauthors{Bond et al.}

\author{J. R. Bond,\altaffilmark{1}
C. R. Contaldi,\altaffilmark{1}
U.-L. Pen,\altaffilmark{1}
D. Pogosyan,\altaffilmark{2}
S. Prunet,\altaffilmark{3}
M. I. Ruetalo,\altaffilmark{1,4}
J. W. Wadsley,\altaffilmark{5}
P. Zhang,\altaffilmark{1,4}
B. S. Mason,\altaffilmark{6,7}
S. T. Myers,\altaffilmark{8}
T. J. Pearson,\altaffilmark{6}
A. C. S. Readhead,\altaffilmark{6}
J. L. Sievers,\altaffilmark{6}
 and
P. S. Udomprasert\altaffilmark{6}}

\altaffiltext{1}{Canadian Institute for Theoretical Astrophysics, 60 St. George Street, Toronto, Ontario M5S 3H8, Canada}

\altaffiltext{2}{Physics Department, University of Alberta, Edmonton, Canada}

\altaffiltext{3}{Institut d'Astrophysique de Paris, 98bis Boulevard Arago, F 75014 Paris, France}

\altaffiltext{4}{Department of Astronomy and Astrophysics, University of Toronto, 60 St. George Street, Toronto, Ontario M5S 3H8, Canada}

\altaffiltext{5}{Department of Physics and Astronomy, McMaster University,
Hamilton, ON L8S 4M1, Canada}

\altaffiltext{6}{California Institute of Technology, 1200 East California Boulevard, Pasadena, CA 91125}

\altaffiltext{7}{National Radio Astronomy Observatory, P.O. Box 2, Green Bank, WV 24944}

\altaffiltext{8}{National Radio Astronomy Observatory, P.O. Box O, Socorro, NM 87801}

\begin{abstract}

We discuss the nature of the possible high-$\ell$ excess in the Cosmic
Microwave Background (CMB) anisotropy power spectrum observed by the
Cosmic Background Imager (\CBI). We probe the angular structure of the
excess in the \CBI\ deep fields and investigate whether it 
could be due to the scattering of CMB photons by hot
electrons within clusters, the Sunyaev-Zeldovich (SZ) effect. We
estimate the density fluctuation parameters for amplitude, $\sigma_8$,
and shape, $\Gamma$, from CMB primary anisotropy data and other
cosmological data. We use the results of two separate hydrodynamical
codes for $\Lambda$CDM cosmologies, consistent with the allowed
$\sigma_8$ and $\Gamma$ values, to quantify the expected contribution
from the SZ effect to the bandpowers of the \CBI\ experiment and  pass
simulated SZ effect maps through our \CBI\ analysis pipeline.  The result is
very sensitive to the value of $\sigma_8$, and is roughly consistent
with the observed power if $\sigma_8 \approx 1$. We conclude that the
\CBI\ anomaly could be a result of the SZ effect for the class of
$\Lambda$CDM concordance models if $\sigma_8$ is in the upper range of
values allowed by current CMB and Large Scale Structure (LSS) data.

\end{abstract}

\keywords{cosmic microwave background --- cosmology: observations}

\section{Introduction}\label{sec:intro}

The Cosmic Background Imager (CBI) provides some of the highest resolution
observations of the cosmic microwave background that have been made
thus far.  The observations cover the multipole range $400<\ell<4000$
which corresponds to collapsed masses ranging from a factor ten larger
than the local group to the largest superclusters.  These observations
show, for the first time, the fluctuations on scales which gave rise
to galaxy clusters and the damping of the power on small scales
\citep{Silk68,Peebles70,Bond87}. Together with the results of other
recent, high precision, CMB experiments
\citep{TOCO,Netterfield02,Lee01,DASI} the observations fit
the scenario of adiabatic fluctuations generated by a period of
inflationary expansion. The \CBI\ observations also provide a unique
insight into angular scales where secondary anisotropy effects are
thought to become an important contribution to the power spectrum.

In a series of papers we have presented the results from the
first year of CMB observations carried out by the \CBI\ between January
and December 2000. A preliminary analysis of the data was presented in
\citet[hereafter Paper~I]{Padin01}. In \citet[hereafter
Paper~II]{Mason02} the observations of three differenced $45^\prime$
(FWHM) deep fields were used to measure the power spectrum to
multipoles $\ell \lesssim 3500$ in wide bands $\Delta\ell \sim
600$. \citet[hereafter Paper~III]{Pearson02} discussed the analysis of
three sets of differenced $145^\prime \times 165^\prime$ mosaiced
fields. Mosaicing gives a telescope a larger effective
primary beam than that defined by the dish radii. This increases
the resolution in the $uv$--plane due to the smaller width of the
convolving function. It also reduces the effect of cosmic variance on
the errors by increasing the sampled area. The mosaic fields provide
high signal--to--noise ratio measurements of the power spectrum up to
multipoles $\ell \sim 1700$ with a resolution $\Delta\ell =
200$. \citet[hereafter Paper~IV]{Myers02} give a detailed description
of our correlation analysis and bandpower estimation methods. These
have enabled us to analyze efficiently the large data sets involved in
the \CBI\ measurements. The implications of the results on cosmological
parameters are described in \citet[hereafter Paper~V]{Sievers02}.

Results from the deep field observations show a fall in the power
spectrum up to $\ell\sim 2000$ which is consistent with the damping
tail due to the finite width of the last scattering surface. Beyond
$\ell\sim 2000$ the power observed is higher than that expected from
standard models of damped adiabatic perturbations which provide
excellent fits to the data at larger scales. An extensive set of tests
have been carried out to rule out possible systematic sources of the
measured excess (Paper~II). Paper III also reports results in this
$\ell$-range which are consistent with those of Paper II.  Since the
thermal noise levels in the \CBI\ mosaic are substantially higher in
this regime than in the deep field observations, the current
discussion will focus on the deep field results.

The deep field spectrum shows the power dropping to
$\ell \sim 2000$, beyond which power levels are significantly higher
than what is expected based on standard models for intrinsic CMB
anisotropy.  Assuming a single bin in flat bandpower above $\ell =
2010$ the observed signal is inconsistent with zero and best--fit
models at the $3.5\sigma$ and $3.1\sigma$ levels respectively. The
one and two sigma confidence intervals for this bandpower are
$359$--$624 \, {\rm \mu K^2}$ and $199$--$946 \, {\rm
\mu K^2}$, respectively, with a best-fit power level of $508 \, {\rm
\mu K^2}$. The confidence limits were obtained by explicitly
calculating the asymmetric, non-Gaussian likelihood of the high-$\ell$
bandpower.

A key consideration in obtaining these results is the
treatment of the discrete radio source foreground.  Sources with
positions known from low-frequency radio surveys were projected out of
the data; the brightest sources were also measured at 31 GHz with the
OVRO 40-m telescope and subtracted directly from the data.  The OVRO
data allowed a complementary treatment of this potentially limiting
population, giving us great confidence in our bright source
treatment. We estimated the power level due to sources too faint to
appear in the low frequency radio source catalogs from 30 GHz number
counts determined from \CBI\ and OVRO data.  The contribution of these
residual sources is $114\, \mu{\rm K}^2$--- a factor of 4.5 below the
observed excess--- and we estimated an uncertainty of $\sim 57\, \mu{\rm
K}^2$ in this contribution. The reader will find further discussion of
both the \CBI\ deep field spectrum and the treatment of the radio
source foreground in Paper II.

Although the significance of the excess power is not conclusive it
provides tantalizing evidence for the presence of secondary
contributions to the microwave background anisotropies. One of the
strongest expected secondary signals is the signature of the
scattering of CMB photons by hot electrons in clusters known as the
Sunyaev-Zeldovich (SZ) effect \citep{SUNYAEV}. The scattering leads to
spectral distortions and anisotropies in the CMB. At the \CBI\
observing frequencies the net effect is a decrement in the temperature
along the line of sight. (However, as described in
\S~\ref{sec:spectrum}, the lead-trail differencing used in the CBI
observations results in positive and negative SZ signals in equal
measure, so this cannot be used as a discriminating signature in our
data.) The SZ effect is expected to dominate over the primary
anisotropies at scales of a few arcminutes.

In this Paper we explore the possibility that the
excess observed in the \CBI\ deep fields may be a signature of the SZ
effect.
In \S~\ref{sec:cosmology} we describe constraints on the
normalization of the mass fluctuations $\sigma_8$ and shape parameter
$\Gamma$ from CMB experiments and a number of independent
surveys. These parameters are critical in determining the amplitude of
the SZ effect over the scales of interest to the \CBI\ results.  In
\S~\ref{sec:analytics} we show the predicted power spectra for the SZ
effect for various cosmological models. We use numerical methods to
estimate the contribution from the SZ effect to the CMB power
spectrum. In \S~\ref{sec:spectrum} we use simulated \CBI\ observations
of SZ `contaminated' CMB realizations to investigate the effect of a
SZ foreground on our bandpower estimation methods. In
\S~\ref{sec:maps} we extend an image filtering technique introduced in
Paper~IV to include specific template filters for the SZ effect.  Our
results and conclusions are discussed in \S~\ref{sec:disc}.
  
\newpage

\section{Amplitude and Shape Parameters for LSS from CMB Data}\label{sec:cosmology}

The SZ power spectrum has a strong dependence on the amplitude of the
density fluctuations. The amplitude is usually parameterized by the rms
of the (linear) mass fluctuations inside 8$h^{-1}$ Mpc spheres,
$\sigma_8$.  We now summarize the constraints we can obtain on the
amplitude and shape of the matter power spectrum from CMB data and
compare them to results from weak lensing surveys and cluster
abundance data. Our parameter estimation pipeline also includes LSS
priors which are designed to encompass the range in estimations from
such experiments, as described below.

We use the parameter $\Gamma$ to define the shape, following
\citet[hereafter BBKS]{BBKS} and \citet[hereafter EBW]{ebw}.  A
byproduct of linear perturbation calculations used to compute ${\cal
C}_\ell$ in our database is transfer functions for density
fluctuations, which can be related to LSS observables. Various
(comoving) wavenumber scales determined by the transport of the many
species of particles present in the universe characterize these
spectra. The most important scale for dark matter dominated universes
is $k^{-1}_{\rm Heq}$, that of the horizon at redshift
$\Omega_{m}/\Omega_{er}$ when the density in nonrelativistic matter,
$\Omega_{m}\bar{a}^{-3}$, equals that in relativistic matter,
$\Omega_{er}\bar{a}^{-4}$. This defines $\Gamma_{eq}$: $k_{\rm Heq}^{-1} =
5 \, \Gamma_{eq}^{-1} \hmpc$, where $\Gamma_{eq} = \Omega_{m} \, {
h} \, [\Omega_{er}/(1.68\Omega_{\gamma})]^{-1/2}$. For the cases we
consider here we simply have fixed the relativistic density to
correspond to the photons and three species of very light neutrinos,
so $\Gamma_{eq} = \Omega_{m} \,  h$.

Certain functional forms for the transfer functions are popular: In
EBW a form was adopted which fit a specific $\Omega_b=0.03$
CDM model, but it is more general to adopt a fit to the $\Omega_b
\rightarrow 0$ form given in BBKS appropriately corrected
for the difference between the temperature of the CMB estimated then
and known so well now \citep{bh95}. Although the coefficients of fits
to detailed models vary with $\omega_b, h, \omega_m$, which in
particular result in oscillations in the transfer function for large
$\omega_b / \omega_m$, it turns out that replacing $\Gamma_{eq}$ by
$\Gamma$ = $\Gamma_{eq}\exp[-(\Omega_b (1+\Omega_{m}^{-1}(2{
h})^{1/2})-0.06)]$ works reasonably well, to about 3\% over the region
most relevant to LSS \citep{sugiyama94supp,bh95}.  Further,
as shown in \citet{yukawa93}, replacing $\Gamma$ by $\Gamma_{\rm eff} =
\Gamma +(n_s-1)/2$ takes into account the main effect of spectral tilt
over this LSS wavenumber band. It has also been shown that the
$\Gamma$-models do fit the APM and 2dFRS data reasonably well at this
stage. Particularly exciting is the prospect that the baryonic
oscillations may be seen, but this is not required by the data
yet. The approximate codification of a vast array of models in a
single $\Gamma_{\rm eff}$ variable simplifies the treatment of LSS in the
CMB data enormously.

The other variable we use to construct the LSS prior is the
combination $\sigma_8 \Omega_{m}^{0.56}$. Although various fits to
cluster abundances give slightly different exponents than the 0.56,
the factor is consistent with a number of other measures and we adopt
this form as a representative value. These invariably involve the
biasing factor $b_g$ for the galaxies involved. For example, relating
the galaxy flow field to the galaxy density field inferred from
redshift surveys takes the form $[b_g\sigma_8] \, \beta_g$, where
$\beta_g = \Omega_m^{0.56}/b_g$ is a numerical factor whose value
depends upon data set and analysis procedure. A combination such as
this also enters into redshift space distortions. A great advantage of
the weak lensing and cluster abundance results shown in the figures is
that they are independent of galaxy bias. However, for the cluster
case, assumptions are needed which, as the spread in estimates
indicates, lead to uncertainties.

\begin{figure}
\plotone{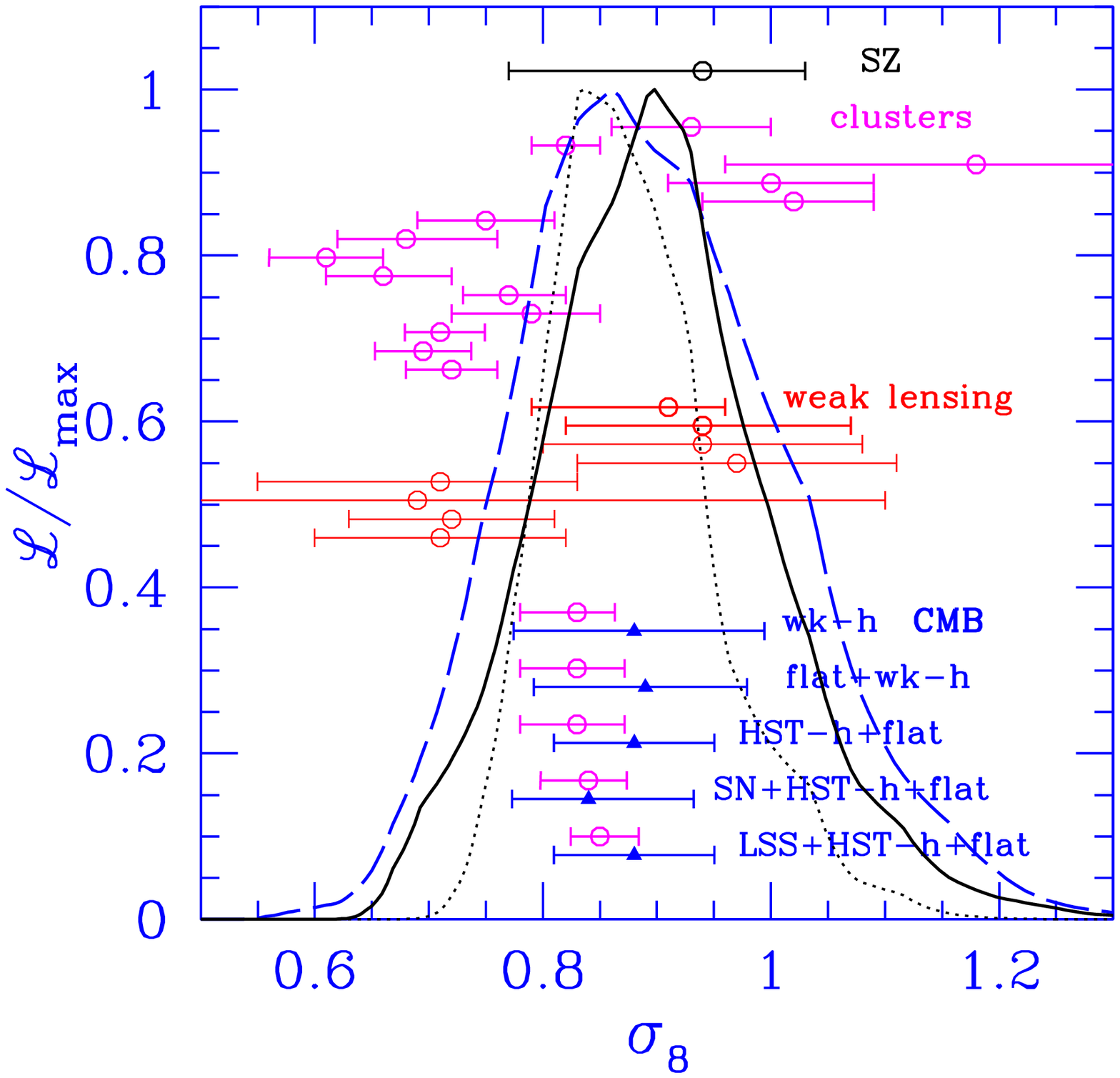}
\caption{One-dimensional projected likelihood functions of
$\sigma_8$ calculated for the CMB data with three prior probability
restrictions on the cosmological parameters are contrasted with
estimates from other datasets. The curves shown used `all-data' for
the CMB available for Paper V, namely \DMR, \DASI\ \citep{DASI},
\Boomerang\ (for the \citet{Ruhl02} cut), \Maxima\ \citep{Lee01}, VSA\
\citep{VSA02} and the \CBI\ mosaic data for the odd $\Delta L$=140
binning. The marginalization has been performed over seven
cosmological variables and all of the relevant calibration and beam
uncertainty variables associated with the experiments, seven in this
case. Application of the weak-h prior is long-dashed blue, of the flat
weak-h prior is solid black and of the LSS flat weak-h prior is dotted
black. Adding \Toco\ \citep{TOCO}, the Boomerang test flight and 17
other experiments predating April 1999 (hereafter `April 99') as well
gives very similar results. The Bayesian 50\% and associated 16\% and
84\% error bars are shown as data points in blue for these and other
priors, in particular those with the stronger HST measurement of $h$
included and with SN data included.  The magenta data points with
smaller errors are those determined with `all-data' to March 2003,
including WMAP, as described in \citet{broysoc03}. The original LSS
prior was constructed based on the cluster abundance data
\citep{bh95,Bond99,Lange01}. An SZ estimate of $\sigma_8$ from
\citet{Goldstein02} that simultaneously determines
amplitudes for a best-fit primordial spectrum and an SZ spectrum for
our \CBI\ deep field data in conjunction with ACBAR \citep{Kuo02} and
BIMA \citep{BIMA02} data is shown at the top. Estimates of $\sigma_8$ from cluster
abundance data and from weak lensing data are also
shown. These results have invariably had more restrictive priors
imposed than those for the CMB, and so are not always applicable, but
the overall level of agreement in the various approaches is
encouraging. From top to bottom, the sample cluster values are 
from
\citet{Eke96,Carlberg97,Fan98,Pen98c,Pierpaoli01,Reiprich01,Seljak01,Viana01,Borgani01},
then two estimates from \citet{Pierpaoli03}, then from
\citet{SBCG03,Allen03,VV03}.  From top to bottom, the weak lensing
estimates are from
\citet{Hoekstra02,Ludo02,Refregier02,Bacon02,Jarvis03,Hamana03,Brown03,Heymans03}. 
\label{fig:priors8} }
\end{figure}

\begin{figure}
\plotone{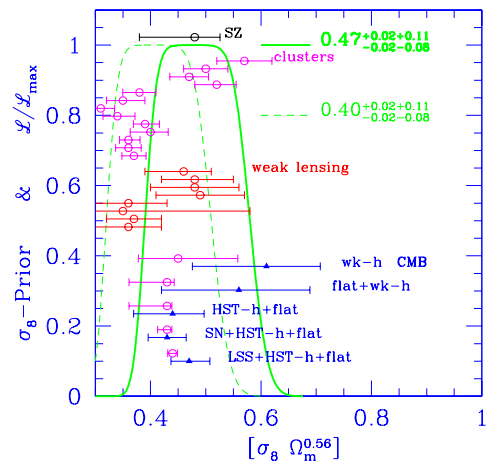} \caption{The weak prior we use for LSS in
    $\sigma_8\Omega_m^{0.56}$ is compared with estimates from SZ,
    cluster abundances, weak lensing and from the CMB data shown in
    Fig.~\ref{fig:priors8}, appropriately scaled. The blue
    error bars are for `all-data' to June 2002, and the magenta are for
    `all-data' to March 2003.  The prior used in past work was shifted
    up in central value from 0.47 to 0.55, but was otherwise the
    same. We have also considered the LSS(low-$\sigma_8$) case (dashed
    green), with the prior shifted downward to be centered on 0.40 to
    accommodate better the low cluster abundance estimates, with
    results shown in Tables~\ref{tab:lss},~\ref{tab:lss1}.}
\label{fig:priors8Om56a}
\end{figure}

In Figures~\ref{fig:priors8} and \ref{fig:priors8Om56a} we show the
effect of the different priors on the distributions of $\sigma_8$ and
$\sigma_8\Omega_m^{0.56}$, respectively. We also compare these to the
weak lensing results shown in the figures. The ``Red Cluster Survey''
results of \citet{Hoekstra02} and the ``Virmos-Descart Survey''
results of \citet{Ludo02} correspond to a version of a weak prior:
they marginalize over $\Gamma_{\rm eff}$ in the range 0.1 to 0.4,
which is largely equivalent for this application to marginalization
over $\omega_b$, $n_s$ and $\omega_m$. They also marginalize over the
uncertainty in the mean redshift of the lensed galaxies $z_s$ from
0.27 to 0.34 in the former case, and 0.78 to 1.08 in the latter
case. On the other hand, the cosmology was kept fixed with $\Omega_m
=0.3$. \citet{Refregier02} and \citet{Bacon02} adopt a more
restrictive parameter range. If the weak marginalization scheme of
\citet{Hoekstra02} and \citet{Ludo02} were used the error bars would
increase in these cases. (Recently \citet{Ludo05} have improved the
Virmos-Descart Survey analysis, which lowers the \citet{Ludo02}
$\sigma_8$ estimate by 12\%.)

Figs.~\ref{fig:priors8} and~\ref{fig:priors8Om56a} show sample cluster
abundance estimates of $\sigma_8$ which are discussed below.

\begin{figure}
\plotone{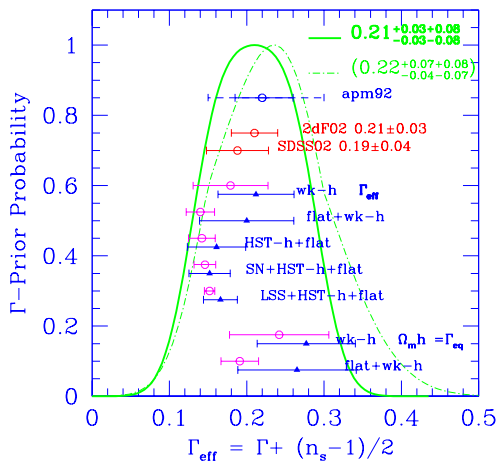} \caption{The prior probability used for the shape
    parameter $\Gamma_{\rm eff}$ is shown as solid green. (This is to be
    contrasted with the more skewed one used in \citet{Lange01}, etc.,
    shown as light dashed-dotted green.) The $\Gamma$-prior was
    designed to encompass the range indicated by the APM data, vintage
    1992, but estimates from the 2dFRS and SDSS shown below are
    quite compatible.  The CMB results for $\Gamma_{\rm eff}$ are
    shown with various choices for priors for comparison. Solid blue
    is for `all-data' as of June 2002, as described in
    \citet{Sievers02}, and magenta is for the data as of March 2003,
    as described in \citet{broysoc03}.  $\Gamma_{\rm eff}$ includes
    corrections for $\omega_b$, $h$, and the tilt.  The related values
    of $\Omega_m h$ are shown at the bottom to show the effect of
    these corrections.  }
\label{fig:priorGam}
\end{figure}

\citet{Ludo02} give a weak lensing result of $\Gamma_{\rm eff} = 0.25 \pm
0.13$ with marginalization over $\Omega_m$ from 0.1 to 0.4 and over
$\sigma_8$. This is not explicitly shown in Fig.~\ref{fig:priorGam}
where we plot the prior probability we adopted for $\Gamma_{\rm eff}$. The APM
result is the long standing one used to construct the original prior,
that $\Gamma_{\rm eff}$ in the 0.15 to 0.3 range provided a good fit to
the data, e.g. EBW and \citet{bh95}. Recent 2dFRS
\citep{Peacock01} and SDSS \citep{Szalay01} results shown give
compatible results.  \citet{Dodelson01} estimate $0.14^{+0.11}_{-0.06}$,
with errors at 95\% confidence, for SDSS. \citet{Szalay01} also give
an estimate of $\sigma_8$ of $0.92 \pm 0.06$, but the issue of galaxy
biasing is folded into this determination. Although there are
indications from the 2dF survey that biasing for the relevant galaxies
is nearly unity from redshift-space distortions ($b=1.04 \pm 0.11$,
\citealt{Verde01}), and the result is therefore compatible with the
results shown, such results are not as directly applicable as those
from clusters and from lensing which directly relate to matter density power
spectrum amplitudes.

The LSS prior used here and in Paper V involves a combination of
constraints on the amplitude parameter $\sigma_8 \Omega_m^{0.56}$ and
on the shape parameter $\Gamma_{\rm eff}$. It differs slightly from that
used in \citet{Lange01}, \citet{Jaffe01} and \citet{Netterfield02}. We
use $\sigma_8 \Omega_m^{0.56}$=$0.47^{+0.02,+0.11}_{-0.02,-0.08}$,
distributed as a Gaussian (first error) smeared by a uniform (tophat)
distribution (second error). We constrain the shape of the power
spectrum via $\Gamma_{\rm eff} \equiv \Gamma + (n_s -
1)/2$=$0.21^{+0.03,+0.08}_{-0.03,-0.08}$, where $\Gamma \approx \Omega_{m}
\, h \, \, e^{-[\Omega_B(1+\Omega_{m}^{-1}(2{ h})^{1/2})-0.06]}$. In earlier
work a central value of $\sigma_8 \Omega_m^{0.56}=0.55$ was
used together with $\Gamma_{\rm eff}$ =
$0.22^{+0.07,+0.08}_{-0.04,-0.07}$. The old $\Gamma$ prior is compared
with the current version in Fig.~\ref{fig:priorGam}. The change does
not affect the parameter values obtained.

The original $\sigma_8$ prior choice was motivated by fits to the
cluster temperature distribution, as was the decision to lower the
central value by 15\% made here. Our philosophy is to make the
distribution broad enough so that reasonable uncertainties are allowed
for in the prior. For example, there are many models that do not fit
the shape as well as the amplitude of the cluster distribution
function. Thus the best $\sigma_8$ for a given model depends upon the
temperature range chosen for the fit, and other physics might be
involved. Especially with the reduced model spaces often considered,
the formal statistical errors can look spectacularly good, but
systematic issues undoubtedly dominate. Curiously, with the 15\% drop,
it appears that the $\sigma_8$ prior chosen could have been designed
for the (weak) weak lensing results. We test sensitivity by showing
results for a `LSS(low-$\sigma_8$)' prior which has a further drop of
15\%: $\sigma_8 \Omega_m^{0.56}$=$0.40^{+0.02,+0.11}_{-0.02,-0.08}$,
This also accommodates some of the recent lower $\sigma_8$ estimates
from cluster abundances, many of which use the X-ray luminosity as an
indicator of mass, calibrated by observations. (We note that the
re-analyzed Virmos-Descart Survey result \citep{Ludo05} gives
$\sigma_8 \Omega_m^{0.56}$=$0.43\pm 0.04$, similar to the value
obtained from the CMB-only data when WMAP is included.)

 We will discuss the simulations in detail below, but we note here
that the cases we have run simulations on have $\sigma_8 =0.9$ and
$1.0$, and $\Gamma_{\rm eff} = 0.18$ and $0.21$.  The detailed
baryonic dependence of the transfer function was included in the
$0.18$ case. The parameter choices were $\omega_b=0.022$ and $0.0245$,
$h=0.7$, $\Omega_m=0.3$ and $0.37$.  A best-fit model to the data for
many prior choices has $\omega_b=0.0225, \omega_c=0.12, n_s=0.95,
\tau_C=0.1$, and $\Omega_\Lambda =0.7$ and with $h=0.69, \Omega_b=0.047,
\Omega_{\rm cdm}=0.253$ with a 13.7 Gyr age; the normalization is
$\sigma_8 =0.84$, and $\Gamma=0.17$ with $\Gamma_{\rm eff} = 0.15$.

In general, CMB data provide weak constraints on the normalization of
the matter fluctuations due to the added effects of the spectral tilt
$n_s$ and the optical depth parameter $\tau_C$ on the overall
amplitude of the CMB power spectrum. Allowing for tensor modes in the
power spectrum introduces further degeneracies. As shown in Paper~V, 
when fitting for a given parameter we adopt the conservative, and more
computationally intensive, approach of marginalizing over all other
variables considered.

\begin{deluxetable*}{cccccccccc}
\tabletypesize{\scriptsize}
\tablecaption{Simulation parameters}
\tablewidth{0pt}
\tablehead{\colhead{Code}
& \colhead{Size (Mpc)}
& \colhead{Resolution}
& \colhead{$\sigma_8$}
& \colhead{$\Omega_bh^2$}
& \colhead{$\Omega_\Lambda$}
& \colhead{$\Omega_m$}
& \colhead{$h$}
& \colhead{$n_s$}
& \colhead{$\Gamma_{\rm eff}$}
}
\startdata
  \SPH
& $200$ 
& $256^3$ 
& $0.9$ 
& $0.0200$ 
& $0.70$
& $0.30$ 
& $0.70$
& $1.0$
& $0.18$
\\
  \SPH
& $200$ 
& $256^3$ 
& $1.0$ 
& $0.0200$ 
& $0.70$
& $0.30$ 
& $0.70$
& $1.0$
& $0.18$
\\
  \SPH
& $400$ 
& $512^3$ 
& $0.9$ 
& $0.0220$ 
& $0.70$
& $0.30$ 
& $0.70$
& $1.0$
& $0.18$
\\
  \MMH
& $143$ 
& $512^3$ 
& $1.0$ 
& $0.0245$ 
& $0.63$
& $0.37$ 
& $0.70$
& $1.0$
& $0.21$
\enddata
\label{tab:simparams}
\end{deluxetable*}

\begin{deluxetable*}{llllll}
\tabletypesize{\scriptsize}
\tablecaption{Amplitude and Shape Parameters for LSS from the CMB:
  `all--data', June 2002 Compilation}
\tablewidth{0pt}
\tablehead{\colhead{Priors}
& \colhead{$\sigma_8$}
& \colhead{$\sigma_8\Omega_m^{0.56}$}
& \colhead{$\Gamma_{eq}$}
& \colhead{$\Gamma$}
& \colhead{$\Gamma_{\rm eff}$}
}
\startdata
wk-$h$
& $0.88^{0.13}_{0.12}$ 
& $0.61^{0.16}_{0.22}$ 
& $0.28^{0.07}_{0.07}$ 
& $0.23^{0.06}_{0.06}$ 
& $0.21^{0.05}_{0.05}$ 
\\
wk-$h$+LSS
& $0.83^{0.11}_{0.10}$ 
& $0.52^{0.06}_{0.11}$ 
& $0.23^{0.04}_{0.04}$ 
& $0.20^{0.04}_{0.04}$ 
& $0.18^{0.03}_{0.03}$ 
\\
wk-$h$+LSS(low-$\sigma_8$)
& $0.81^{0.10}_{0.10}$ 
& $0.45^{0.06}_{0.08}$ 
& $0.21^{0.04}_{0.04}$ 
& $0.17^{0.04}_{0.04}$ 
& $0.17^{0.03}_{0.03}$ 
\\
wk-$h$+SN
& $0.85^{0.15}_{0.12}$ 
& $0.45^{0.11}_{0.08}$ 
& $0.21^{0.03}_{0.03}$ 
& $0.17^{0.03}_{0.03}$ 
& $0.17^{0.03}_{0.03}$ 
\\
\\[0.01cm]\tableline\\[0.01cm]
flat+wk-$h$
& $0.89^{0.10}_{0.11}$ 
& $0.56^{0.23}_{0.25}$ 
& $0.26^{0.08}_{0.08}$ 
& $0.23^{0.08}_{0.08}$ 
& $0.20^{0.07}_{0.07}$ 
\\
flat+wk-$h$+LSS
& $0.87^{0.08}_{0.07}$ 
& $0.50^{0.07}_{0.11}$ 
& $0.23^{0.04}_{0.04}$ 
& $0.20^{0.04}_{0.04}$ 
& $0.17^{0.03}_{0.03}$ 
\\
flat+wk-$h$+LSS(low-$\sigma_8$)
& $0.84^{0.09}_{0.08}$ 
& $0.43^{0.06}_{0.09}$ 
& $0.20^{0.04}_{0.04}$ 
& $0.17^{0.03}_{0.03}$ 
& $0.15^{0.03}_{0.03}$ 
\\
flat+wk-$h$+SN
& $0.85^{0.11}_{0.09}$ 
& $0.43^{0.10}_{0.11}$ 
& $0.20^{0.03}_{0.03}$ 
& $0.17^{0.03}_{0.03}$ 
& $0.15^{0.03}_{0.03}$ 
\\
\\[0.01cm]\tableline\\[0.01cm]
flat+HST-$h$
& $0.86^{0.11}_{0.11}$ 
& $0.44^{0.13}_{0.16}$ 
& $0.21^{0.05}_{0.05}$ 
& $0.18^{0.05}_{0.05}$ 
& $0.16^{0.04}_{0.04}$ 
\\
flat+HST-$h$+LSS
& $0.88^{0.08}_{0.08}$ 
& $0.47^{0.08}_{0.07}$ 
& $0.22^{0.03}_{0.03}$ 
& $0.19^{0.03}_{0.03}$ 
& $0.17^{0.03}_{0.03}$ 
\\
flat+HST-$h$+SN
& $0.84^{0.11}_{0.08}$ 
& $0.43^{0.08}_{0.08}$ 
& $0.20^{0.03}_{0.03}$ 
& $0.17^{0.03}_{0.03}$ 
& $0.15^{0.03}_{0.03}$ 
\\
flat+HST-$h$+LSS+SN
& $0.87^{0.09}_{0.08}$ 
& $0.45^{0.07}_{0.05}$ 
& $0.21^{0.02}_{0.02}$ 
& $0.18^{0.02}_{0.02}$ 
& $0.16^{0.03}_{0.03}$ 
\\
flat+HST-$h$+LSS(low-$\sigma_8$)+SN
& $0.85^{0.09}_{0.07}$ 
& $0.43^{0.05}_{0.05}$ 
& $0.20^{0.02}_{0.02}$ 
& $0.17^{0.02}_{0.02}$ 
& $0.15^{0.02}_{0.02}$ 
\\
\enddata
\tablecomments{Amplitude and shape parameter estimates from various
datasets, for various prior probability choices and using
`all-data' for the CMB available for Paper V, and as described
there, in particular DMR+\DASI+\Maxima\ +
\Boomerang\ (Ruhl et al. 2003 cut), VSA and \CBI\ mosaic data
for the odd $\Delta \ell$=140 binning, as described in Paper
III. In the first four rows, the weak prior in $h$ ($0.45 < h <
0.90$) is imposed (including further weak constraints on cosmological
age, $t_0>10$~Gyr, and matter density, $\Omega_m>0.1$). The sequence
shows what happens when priors for LSS (with $\sigma_8\Omega_m^{0.56}$ centered
about 0.47), for LSS(low-$\sigma_8$) (with the $\sigma_8$ distribution
shifted downward by 15\%, centered about 0.40), and for SN are
imposed. While the first four rows allow $\Omega_{tot}$ to be free,
the next four have $\Omega_{tot}$ pegged to unity, a number strongly
suggested by the CMB data.  The final 4 rows show the `strong-$h$'
prior, a Gaussian centered on $h=0.71$ with dispersion $\pm 0.07$,
obtained for the Hubble key project. The $t_0>10$~Gyr and
$\Omega_m>0.1$ constraints are also imposed, but they have no
impact. Central values and $1\sigma$ limits for the 7 database
parameters that form our fiducial minimal-inflation model set are
found from the 16\%, 50\% and 84\% integrals of the marginalized
likelihood for $\sigma_8$ and $\sigma_8\Omega_m^{0.56}$. For
$\Gamma_{eq} \equiv \Omega_m h$, $\Gamma$ and $\Gamma_{\rm eff}$, the
values are means and variances of the variables calculated over the
full probability distribution.  Shifting the center of the $\sigma_8
\Omega_m^{0.56}$ prior downward by 15\% to accommodate more of the low
cluster results has little effect on the flat+HST-$h$+LSS+SN result.}
\label{tab:lss}
\end{deluxetable*}

In Table~\ref{tab:lss} we summarize the constraints on the LSS
parameters from the CMB data used in Paper~V and a combination of
priors. The mean and quoted ($1\sigma$) errors correspond to the 50\%,
16\% and 84\% integrals of the marginalized likelihoods
respectively. 
\begin{figure}
\plotone{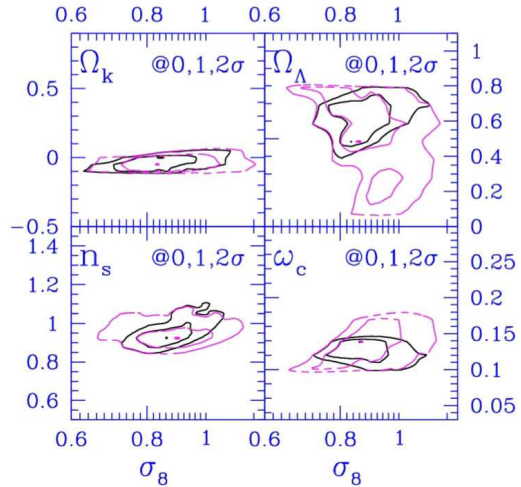} \caption{One and two sigma contours of 2D projected
    likelihood functions show how various cosmological parameters
    correlate with $\sigma_8$. For this case, `all--data' from the
    June 2002 compilation was used: \DASI+\Maxima+\Boomerang+VSA plus
    the \CBI\ mosaic data for the odd $\Delta L$=140 binning. The
    $\Omega_k-\sigma_8$ panel shows the weak-$h$ prior (magenta) and
    LSS+weak-$h$ prior (solid black).  For the other 3 panels the flat
    constraint was added to these two priors as well. Note the
    positive correlation with $n_s$, but little correlation in the other
    variables.}
\label{fig:s8}
\end{figure}

\begin{figure}
\plotone{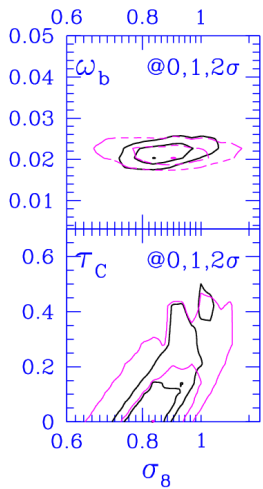} \caption{One and two sigma contours of 2D projected
    likelihood functions for $\omega_b$ and $\tau_C$ for `all-data' of
    \citet{Sievers02}. The flat+weak-$h$ prior (magenta) and
    flat+LSS+weak-$h$ prior (solid black) cases are shown. The
    significant $\sigma_8$--$\tau_C$ correlation is evident, which
    results in a higher $\sigma_8$ for higher $\tau_C$. One could
    impose a stronger prior than $\tau_C < 0.7$ (as is done here)
    based on astrophysical arguments. This is fraught with
    uncertainty since it involves the first objects collapsing on
    small scales in the universe and their efficiency in generating
    stars that produce ionizing radiation. However, $\tau_C$
    has apparently been detected by WMAP at the $0.16 \pm 0.04$ level
    \citep{Kogut03}, and results with a prior encompassing this detection
    on the March 2003 compilation of the data are given in
    Table~\ref{tab:lss1}.}
\label{fig:s8tau}
\end{figure}

In Figs.~\ref{fig:s8} and~\ref{fig:s8tau} we show the correlation of
$\sigma_8$ with other parameters considered.  Correlations with
$\tau_C$ are nontrivial, and allowing the space to be open in $\tau_C$
does tend to give higher marginalized $\sigma_8$. The CMB data as of
June 2002 did not give an indication of a $\tau_C$ detection, and we
have imposed no physical prior on those results. 

\begin{deluxetable*}{llllll}
\tabletypesize{\scriptsize} 
\tablecaption{Amplitude and Shape
Parameters for LSS from the CMB: 'all--data', March 2003 Compilation}
\tablewidth{0pt} 
\tablehead{\colhead{Priors} & \colhead{$\sigma_8$} &
\colhead{$\sigma_8\Omega_m^{0.56}$} & \colhead{$\Gamma_{eq}$} &
\colhead{$\Gamma$} & \colhead{$\Gamma_{\rm eff}$} } 
\startdata 
wk-$h$
& $0.83^{0.04}_{0.06}$ 
& $0.45^{0.24}_{0.16}$ 
& $0.24^{0.07}_{0.07}$ 
& $0.20^{0.06}_{0.06}$ 
& $0.18^{0.05}_{0.05}$ 
\\ wk-$h$+LSS 
& $0.85^{0.04}_{0.03}$ 
& $0.44^{0.02}_{0.02}$ 
& $0.21^{0.02}_{0.02}$ 
& $0.17^{0.01}_{0.01}$ 
& $0.15^{0.02}_{0.02}$ 
\\
wk-$h$+LSS(low-$\sigma_8$)
& $0.84^{0.05}_{0.05}$ 
& $0.43^{0.02}_{0.04}$ 
& $0.20^{0.02}_{0.02}$ 
& $0.17^{0.02}_{0.02}$ 
& $0.15^{0.02}_{0.02}$ 
\\ 
wk-$h$+SN 
& $0.84^{0.05}_{0.05}$ 
& $0.43^{0.02}_{0.13}$ 
& $0.20^{0.02}_{0.02}$ 
& $0.17^{0.02}_{0.02}$ 
& $0.14^{0.02}_{0.02}$ 
\\
\\[0.01cm]\tableline
\\[0.01cm] 
flat+wk-$h$ 
& $0.83^{0.05}_{0.06}$ 
& $0.43^{0.03}_{0.16}$ 
& $0.19^{0.03}_{0.03}$ 
& $0.16^{0.03}_{0.03}$ 
& $0.14^{0.02}_{0.02}$ 
\\ 
flat+wk-$h$+LSS 
& $0.85^{0.04}_{0.03}$ 
& $0.44^{0.02}_{0.02}$ 
& $0.21^{0.01}_{0.01}$ 
& $0.17^{0.01}_{0.01}$ 
& $0.15^{0.01}_{0.01}$ 
\\
flat+wk-$h$+LSS(low-$\sigma_8$)
& $0.84^{0.05}_{0.05}$ 
& $0.43^{0.02}_{0.04}$ 
& $0.20^{0.02}_{0.02}$ 
& $0.17^{0.02}_{0.02}$ 
& $0.15^{0.02}_{0.02}$ 
\\ 
flat+wk-$h$+SN 
& $0.84^{0.05}_{0.05}$ 
& $0.43^{0.02}_{0.13}$ 
& $0.20^{0.02}_{0.02}$ 
& $0.17^{0.02}_{0.02}$ 
& $0.14^{0.02}_{0.02}$ 
\\ 
\\[0.01cm]\tableline
\\[0.01cm] 
flat+HST-$h$ 
& $0.83^{0.05}_{0.06}$ 
& $0.43^{0.02}_{0.16}$ 
& $0.19^{0.03}_{0.03}$ 
& $0.16^{0.03}_{0.03}$ 
& $0.14^{0.02}_{0.02}$ 
\\ flat+HST-$h$+LSS 
& $0.85^{0.04}_{0.03}$ 
& $0.44^{0.02}_{0.02}$ 
& $0.21^{0.01}_{0.01}$ 
& $0.17^{0.01}_{0.01}$ 
& $0.15^{0.01}_{0.01}$ 
\\ flat+HST-$h$+SN 
& $0.84^{0.04}_{0.05}$ 
& $0.43^{0.02}_{0.04}$ 
& $0.20^{0.02}_{0.02}$ 
& $0.17^{0.02}_{0.02}$ 
& $0.15^{0.02}_{0.02}$ 
\\ flat+HST-$h$+LSS+SN 
& $0.85^{0.04}_{0.03}$ 
& $0.44^{0.02}_{0.02}$ 
& $0.21^{0.01}_{0.01}$ 
& $0.17^{0.01}_{0.01}$ 
& $0.15^{0.01}_{0.01}$ 
\\ 
flat+HST-$h$+LSS(low-$\sigma_8$)+SN
& $0.85^{0.04}_{0.04}$ 
& $0.44^{0.02}_{0.03}$ 
& $0.20^{0.02}_{0.02}$ 
& $0.17^{0.02}_{0.02}$ 
& $0.15^{0.02}_{0.02}$ 
\\ 
\enddata
\tablecomments{Similar to Table.~\ref{tab:lss} but for an extension
of `all-data' to include WMAP, ACBAR, extended VSA (\citep{VSAext02}) and \CBI\
results, as well as DASI, Boomerang and Maxima, analyzed in the same
way. A further (weak) prior on $\tau_C$ was applied to accommodate the
WMAP detection from the cross-correlation of temperature and
polarization, but this makes little difference to the results.  For
example, without this extra prior, flat+wk-$h$ shifts to
$0.81^{0.05}_{0.05}$ and $0.41^{0.03}_{0.11}$. As for
Table~\ref{tab:lss}, the downshifted LSS prior has little effect on
flat+HST-$h$+LSS+SN. }
\label{tab:lss1}
\end{deluxetable*}

The addition of the first-year WMAP data and other data that has
appeared since sharpens the determination of the parameters
$\sigma_8$, $\sigma_8 \Omega_m^{0.56}$ and $\Gamma_{\rm eff}$. This `March
2003' compilation of CMB anisotropy data, described in \citet{broysoc03},
is treated in the same way as the June 2002 data is in Paper V, with
the same ${\cal C}_\ell$-database. Results are shown in
Table~\ref{tab:lss1} and in the reduced errors in
Figs.~\ref{fig:priors8}, \ref{fig:priors8Om56a}
and~\ref{fig:priorGam}. For the March 2003 data, a prior was included
on $\tau_C$ to accommodate the $\tau_C =0.16 \pm 0.04$ detection 
reported by the WMAP team \citep{Kogut03}. The form we have adopted
has a top-hat spread convolved with a Gaussian distribution, as for
our $\sigma_8 \Omega_m^{0.56}$ and $\Gamma_{\rm eff}$ priors: $\tau_C =
0.16^{+0.04,+0.06}_{+0.04,+0.06}$. Similar results are obtained using
the Markov Chain Monte Carlo approach, which explicitly includes the
$\tau_C$-detection using the TE data of \citet{Kogut03}.


Best-fit models may have lower $\sigma_8$ values than the
marginalized values. Typical best--fit models for the June 2002
dataset are: ($\Omega_{\rm cdm}=0.47, \Omega_{\Lambda}=0.50,
\Omega_b=0.079, h=0.51, n_s=0.90, \tau_C=0$ and $\sigma_8=0.72$) for
the weak prior case and ($\Omega_{\rm cdm}=0.44,
\Omega_{\Lambda}=0.50, \Omega_b=0.063, h=0.57, n_s=0.90, \tau_C=0$ and
$\sigma_8=0.81$) for flat+weak and flat+weak+LSS priors. These
$\sigma_8$ values are lower because the best--fits selected the
$\tau_C=0$ peak in the likelihood whereas integration over the
relatively broad $\tau_C$ likelihood allows for the inclusion of
high--$\sigma_8$ models. If the HST-$h$ or SN prior is included along
with LSS, the best-fit model is a more conventional $\Lambda$CDM one,
with $\tau_C=0.1$: ($\Omega_{\rm cdm}=0.25, \Omega_{\Lambda}=0.70,
\Omega_b=0.047, h=0.69, n_s=0.95, \tau_C=0.1$ and
$\sigma_8=0.84$). For the March 2003 data with the $\tau_C$ prior,
this $\tau_C=0.1$ model also provides a best-fit for the weak+flat and
weak+flat+LSS cases.

Comparison of our results with independent estimates of $\sigma_8$
from CMB data is complicated by the different choices of parameter
marginalization and by different treatments of the CMB
data. \citet{Lahav02} carried out a joint CMB--2dFRS analysis of
cosmological parameters. Their result is $\sigma_8=0.73 \pm 0.05$
after marginalizing over some variables but keeping $\tau_C$ fixed at
0 and $n_s$ fixed at unity. This value is therefore biased low with
respect to non--zero $\tau_C$ models.  They also show how letting
$\tau_C$ vary gives higher values for $\sigma_8$. \citet{Melchiorri02}
quote smaller values for $\sigma_8$ in their analysis, even when
$\tau_C$ was allowed to vary. 

The values we obtain for the March 2003 dataset are in good agreement
with those obtained by the WMAP team \citep{Spergel03}. However we
caution that the projected distribution of $\sigma_8$ from the CMB
depends upon the other parameters. For the seven-parameter
inflation-motivated models considered here, there is a strong
correlation of $\sigma_8$ with the Thomson scattering depth $\tau_C$
and the primordial spectral index $n_s$. If, in addition, $n_s$ is
allowed a logarithmic correction with wavenumber, introducing another
parameter highly correlated with $\sigma_8$, the distribution extends
to higher $\sigma_8$ \citep{broysoc03}. Allowing for a gravity wave
contribution extends the distribution to lower $\sigma_8$. These
caveats about the effect of adding extra parameters in the CMB
analysis together with the scatter in the estimates of $\sigma_8$ from
clusters and weak lensing shown in Figs.~\ref{fig:priors8} and
\ref{fig:priors8Om56a} reflect the current uncertainty as to whether a
low or high $\sigma_8$ will emerge.

\section{The Angular Power Spectrum of the Sunyaev-Zeldovich Effect}\label{sec:analytics}

The SZ effect is a signature of the scattering of CMB photons off hot
electrons. The effect can be described in
terms of the fractional energy gain per scatter along the line of
sight. By multiplying by the number density of electrons and
integrating along the line of sight we can derive the induced 
temperature change:
\begin{equation}
\frac{\Delta T_{\rm SZ}}{T_{\rm CMB}} = -2 y \, \psi_K(x)=
-2\sigma_{T}\int n_e\frac{k_B(T_e-T_{\rm CMB}) }{m_ec^2}d\chi \, \psi_K(x)\, , 
\end{equation}
where $y$ is the Compton $y$-parameter, $\sigma_T$ is the
Thomson cross--section, $n_e$ is the electron number density, $T_e$ is
the electron temperature, $T_{\rm CMB}$ is the CMB temperature, $k_B$ is
Boltzmann's constant and $m_ec^2$ is the electron's rest mass
energy. Here $\psi_K(x), \, x\equiv h\nu /k_B T_{\rm CMB}$ is a
frequency-dependent function which is unity at Rayleigh-Jeans
wavelengths and is 0.975 at the 30 GHz frequencies probed by the \CBI.

With the great increase in experimental sensitivity, measurement of
the SZE in known clusters has become almost routine (see, e.g.,
\citealt{Birkinshaw84,Carlstrom96,Holzapfel97,Mason01,Udomprasert01,Grainge02}). Coupled
with X--ray observations of the cluster these measurements yield
independent constraints on the value of the Hubble constant. They also
provide a direct measurement of the amount of baryon mass in the
cluster gas. The SZ effect is expected to contribute significantly to
the CMB power spectrum at scales $\ell > 2000$, with a crossover point
between the primary CMB and SZ signature occuring somewhere between
$\ell\sim 2000$ and $\ell\sim 3000$. Surveys observing at these scales
will therefore require accurate component separation to reconstruct
the primary CMB spectrum and much work has focused on this issue in
recent years. Many proposed experiments will adopt multi-frequency
observing strategies in order to separate the different components by
their different spectral dependence. These techniques cannot be
applied to the narrow frequency band of the \CBI\
observations. However we can attempt to address the question of
whether the SZ effect could provide a contribution with the required
amplitude to explain the observed excess.

We use the output of two separate hydrodynamical simulation algorithms
in an attempt to predict the level at which the SZ effect will
contribute to the \CBI\ deep field observations. We also relate  our
numerical results to analytical models of the SZ power spectrum
based on the halo model. 

\subsection{Hydrodynamical Simulations of the Sunyaev-Zeldovich Effect}

The two codes that we have employed both provide high-resolution dark
matter and hydrodynamical simulations of large scale cosmic structure which we
use to generate simulated wide-field SZ maps. The simulation
algorithms and processing techniques were developed independently
\citep{Pen98a,Bond98,Bond01,Wadsley03} and are based on two
different numerical schemes for solving the self--gravitating,
hydrodynamical equations of motion. High-resolution gasdynamical
simulations of large enough volumes are still beyond current
technological capabilities. Following an approach taken previously by
other authors \citep[e.g.,][]{Springel01}, we create a
pseudo-realization of the cosmic structure up to high redshifts from a
single, medium-sized, high-resolution simulation by stacking randomly
translated and rotated (or flipped) copies of the (evolving) periodic
volume.

One set of SZ simulations was obtained using the \textsc{Gasoline}
code, an efficient implementation of the (Lagrangian) smooth particle
hydrodynamics (\SPH) method \citep{Wadsley03}. This tree$+$\SPH\
code uses a pure tree-based gravity solver and has spatially-adaptive
time stepping. It has been parallelized on many architectures (MPI in
this case) and shows excellent scalability. The results presented here
are based on the analysis of three high resolution $\Lambda$CDM
simulations: two 200 Mpc box computations, with $256^3$ dark matter
plus $256^3$ gas particles, and one 400 Mpc box computation, with
$512^3$ dark and $512^3$ gas particles, a very large number for SPH
simulations. The calculations were adiabatic, in the sense that only
shocks could inject entropy into the medium. Despite the different
sizes, all three simulations were therefore run with the same mass
resolution. All simulations were run with a gravitational softening of
50 kpc (physical). The scale probed by the \SPH\ smoothing kernel was
not allowed to become smaller than the gravitational softening in the
200 Mpc runs, but it was not limited in the 400 Mpc run, which
attained gas resolution scales as low as 5 kpc in the highest density
environments, though increasing considerably at lower density. (64
neighbors were required to be within the smoothing kernel in the 400
Mpc run.) The $512^3$ simulation was performed on a large-memory, 114
(667 MHz) processor \textsc{SHARCNET COMPAQ SC} cluster at McMaster
University. It required about 80 GB of memory and took $\sim 40$ days
of wall time to run. 

Another set of SZ maps was obtained using a $512^3$ run of the Moving
Mesh Hydrodynamics (\MMH) code of \citet{Pen98a}. This code features a
full curvilinear Total-Variation-Diminishing (TVD) hydro code with a
curvilinear particle mesh (PM) N-body code on a moving coordinate
system. We follow the mass field such that the mass per unit grid cell
remains approximately constant.  This gives all the dynamic range
advantages of \SPH\ simulations combined with the speed and high
resolution of grid algorithms. The box size was $143$ Mpc and the
smallest grid spacing was $57$ kpc (comoving). This $512^3$ simulation
used $30$ GB memory and took about three weeks ($\sim 1500$ steps) on
a 32 processor shared memory Alpha GS320 using Open MP parallelization
directives. The calculations were adiabatic as well. 

The simulation boxes yield a number of projections of the gas
distributions in random orientations and directions. The projections
are then stacked to create a redshift range appropriate for the SZ
simulation. In the case of the \MMH\ simulation the angular size of
the simulated box above a redshift of $z=1.6$ is smaller than $2$
degrees, the required angular size of the SZ simulations. To include
the contributions from higher redshifts we tiled two copies of the
periodic box to cover the increased angles. This only affects the
smallest scales. All the maps were obtained with a low redshift cutoff
at $z=0.2$ to minimize the contribution from the closest clusters in
the projections. The computational costs of running such large
hydrodynamical simulations prevented us from obtaining targeted
simulations with identical cosmological parameters from both
algorithms. Below we account for the parameter variations in the
models run when comparing the two simulations and we see a remarkable
consistency between the results. The parameter variations are, in
fact, an advantage for us since they provide us with a sampling of the
SZE amplitude for processing through the CBI pipeline. We summarize
the parameters used in the simulations in Table~\ref{tab:simparams}.

The two 200 Mpc \SPH\ simulations were used to generate 20 2$\times$2 degree
maps each while the \MMH\ simulation yielded 40 separate maps. A
detailed analysis of the \MMH\ maps including SZ statistics and
non-Gaussianity is given by \citet{Zhang02}. A preliminary analysis of
simulated SZ maps based on one of the $256^3$ \SPH\ simulations was
presented by \citet{Bond01}, and a more detailed analysis of the
$512^3$ results will be presented elsewhere.

\begin{figure}
\plotone{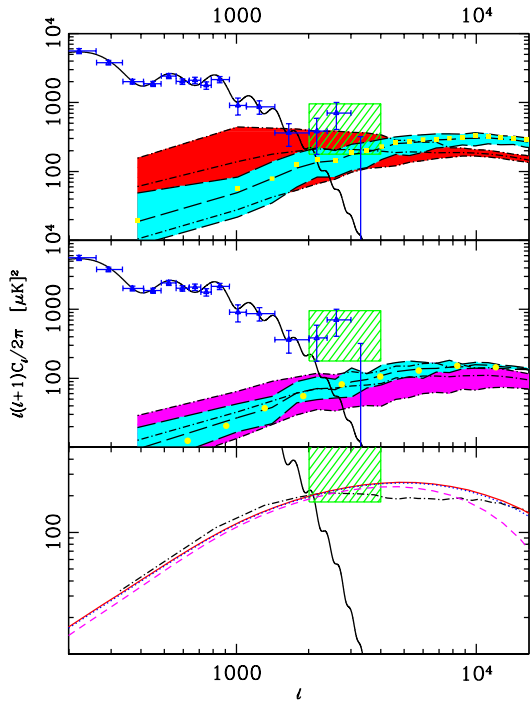}
\caption{SZ power spectra for various simulations. All SZ
spectra have been scaled to a common $\Omega_bh=0.0314$, the best--fit
to the CMB data. The upper panel shows the SZ spectra for the
$\sigma_8=1.0$ \SPH\ $256^3$ 200 Mpc (cyan) and \MMH\ $512^3$ 140 Mpc
(red) simulations.  We also plot the power spectrum from the
$\sigma_8=0.9$ \SPH\ $256^3$ 200 Mpc run (yellow points) scaled using
the relation of equation~\ref{eq:scaling}. The target 2-sigma region
suggested by the deep data is shown as a green-hatched box. An optimal
power spectrum combining all \CBI\ data with the \Boomerang, \COBE,
\DASI\ and \Maxima\ data is shown as blue triangles. The solid black
curve is a best-fit model to the data out to $\ell =2000$. The middle
panel compares the $\sigma_8=0.9$ $512^3$ 400 Mpc (cyan) and $256^3$
200 Mpc (magenta) \SPH\ simulations. The yellow points are the spectra
derived from the $\Lambda$CDM simulation of \citet{Springel01}, also
scaled to $\Omega_bh=0.0314$. The bottom panel compares the \MMH\
results (dot-dash) with the analytic halo model results described in
\S~\ref{sec:analytics} for mass cuts of $0.001M_8$ (solid, red),
$0.01M_8$ (dotted, blue) and $0.1M_8$ (dashed, magenta) described in
the text.  There is little sensitivity to a change in the lattice size
over the $\ell$-range shown. We note that physical effects, such as early
entropy injection, may change
the effective mass cut. 
\label{fig:szspectra}  }
\end{figure}

\subsection{Hydrodynamical Simulation Results}

In Fig.~\ref{fig:szspectra} we show the results for the
 $\sigma_8=1.0$, $143$ Mpc \MMH\ simulation, the $\sigma_8=1.0$ and
 $0.9$, 200 Mpc \SPH\ simulation and $\sigma_8=1.0$, 400 Mpc \SPH\
 simulation. The curves show averages for the ${\cal C}_{\ell}$
 spectra from the 40 and 20 \MMH\ and \SPH\ maps respectively. The
 shaded regions show the full excursion of the power spectra in each
 set of maps. This gives a partial indication of the scatter induced
 in the power due to the small areas being considered. We compare
 these with a best--fit model to the \Boomerang, \CBI, \COBE, \DASI\ and
 \Maxima\ data and an optimal combination of the mosaic and deep
 \CBI, \Boomerang, \COBE, \DASI\ and \Maxima\ bandpowers (Paper~V). The
 combined spectrum is designed for optimal coverage with variable
 bandwidths over the range of scales considered. At scales above
 $\ell\sim 1000$ the optimal spectrum is dominated by the
 contributions of the \CBI\ mosaic and deep results.

To compare the cosmological simulations we have to account for the
difference in parameters the simulations were run with. The dominant
effect for the SZ ${\cal C}_{\ell}$ spectrum is variation in
$\sigma_8$ and $\Omega_bh$, scaling as $\sigma_8^7$ and
$(\Omega_bh)^2$. The top panel of Fig.~\ref{fig:szspectra}
demonstrates that ${\cal C}_{\ell}\propto \sigma_8^7$ does indeed
bring the 200 Mpc \SPH\ run at $\sigma_8=0.9 $ into essentially perfect
alignment with the $\sigma_8=1$ run (yellow/square points). (Apart
from the amplitude, the initial conditions in the simulations were
otherwise the same). The exact factor $n$ in a scaling of form ${\cal
C}_{\ell}\sim (\Omega_bh)^2\sigma_8^{n}$ depends upon the model in
question, but $6 \lta n \lta 9$ is typical
\citep{Zhang02}. Differences in $\Omega_m$ and shape of the power
spectrum also have an influence on ${\cal C}_{\ell}$, as
\citet{Komatsu:2002wc} discuss in more detail.

For the two separate $\sigma_8$ cases employed in the runs we rescale
the spectra to a nominal $\Omega_bh = 0.0314$, a value suggested by
the CMB data. This enables us to compare the \MMH\ result with the
\SPH\ results for the $\sigma_8=1$ runs (top panel
Fig.~\ref{fig:szspectra}). We see that the two codes give similar
amplitudes over the range considered, and, in particular, these are
very similar over the scales of interest for the \CBI\ deep field
result (green/hatched box). At larger angular scales, the \MMH\
simulation shows a somewhat higher amplitude than the \SPH\
simulation. This may be in part due to the presence of a large rare
cluster in the \MMH\ volume simulation. The Poisson noise contribution
of the large cluster is aggravated by the resampling technique adopted
in the SZ map making stage described above. Variable redshift cutoffs
do indeed confirm the dominant effect of this single cluster on the
low--$\ell$ tail of the spectrum.

At smaller angular scales the two results start to diverge. The
differences in spectral shape and cosmological parameters should not
account for this. It may be attributable to the different techniques
used to limit the resolution achievable, but this needs more
investigation.

In the middle panel we have plotted the results from the 400 Mpc and
200Mpc \SPH\ simulations which were both run with $\sigma_8=0.9$. We
also compare this run to SZ power spectra derived from the 200 Mpc box
\SPH\ simulations of \citet{Springel01}. The cosmological parameters
of the \citet{Springel01} simulations were the same as those used in
our \SPH\ runs except for a value of $\Omega_bh^2=0.018$. We also
rescale this spectrum to the nominal $\Omega_bh=0.0314$. The agreement
is remarkable. We can contrast this level of agreement with the
situation described in \citet{Springel01} where it appeared that
different codes were giving quite different results. Nonetheless we
plan a further exploration of the effects of lattice size variations
and other numerical parameters to compare more exactly the \MMH\ and
\SPH\ codes.

Based on these results we can calibrate the expected
power to compare with the wide band result of Paper~II:
\begin{eqnarray}
        \mbox{SPH:} && {\cal C}^{SZ}_{\ell} \sim 170\, \mu K^2
        \frac{(\Omega_bh)^2}{(0.0314)^2}\sigma_8^7\nonumber\\
        \mbox{MMH:} && {\cal C}^{SZ}_{\ell} \sim 210\, \mu K^2
        \frac{(\Omega_bh)^2}{(0.0314)^2}\sigma_8^7.
\label{eq:scaling}      
\end{eqnarray}
For $\sigma_8=1$ and $\Omega_bh=0.0314$, these amplitudes are $\sim
0.4$ relative to the noise level for the CBI joint analysis of the
three deep fields. 

\subsection{Analytic Modeling of the Sunyaev-Zeldovich Power
Spectrum}

The general analytic framework comes under the generic name of `halo
models', in which simple parameterized gas profiles within clusters
and groups are constructed, appropriately scaled according to the
masses. The gas profile in a halo and halo mass-temperature relation
determines the SZ effect of the halo. The abundance of the halos as a
function of mass and redshift is determined by the \citet{Press74}
formula, the \citet{BM96} peak-patch formula, or formulas derived from fits to
$N$-body simulations. Clustering of the halos is included through
simple linear biasing models.  The halo model approach has been
applied to the SZ effect by many authors over time (e.g.,
\citealt{Bond88,Cole88,Makino93,BM96,Atrio99,Komatsu99,Cooray00,Molnar00,Seljak01b}).
There has still not been an adequate calibration of the relation
between the physical parameters describing the gas distribution in the
halos and the results of gas dynamical simulations, and usually the
parameters that have been adopted are those derived from observations
of clusters at low redshift. We show here that we can get reasonable
fits to our simulation results by choosing suitable gas profile
parameters. However we caution that the gas profile is a function of
halo mass and redshift, so globally fitting to our simulated SZ power
spectra may not deliver reliable parameters for the analytic
model. For now, we take our globally fit models to show sensitivity
to parameter variations.

In the bottom panel of Fig.~\ref{fig:szspectra} a few analytic models
are shown to compare with the $\sigma_8=1$ \MMH\ simulation. The
Press-Schechter distribution for the halo comoving number density $n$
as a function of halo mass $M$ and redshift $z$ is $Mdn/dM$ $ \propto
$ $({2}/{\pi})^{1/2}$ $(\bar{\rho}_0/M) \left| {d\ln\sigma}/{d\ln M}
\right| \nu e^{-\nu^2/2}$, where $\nu \equiv
(\delta_{c}/{\sigma})$. Here $\bar{\rho}_{0}$ is the present mean
matter density of the universe. $\sigma(M,z)$ is the linear theory
{\it rms} density fluctuation in a sphere containing mass $M$ at
redshift $z$.  It is calculated using the input linear density power
spectrum $P(k)$ of our simulations, which is truncated at $k_{\rm
lower}=2\pi/L$ and $k_{\rm upper}=\sqrt{3}\pi N/L$ due to the finite
box size $L$ and resolution $N$ of simulations, respectively. In this
formula, we have taken $\delta_{c}=1.686$ as the linearly extrapolated
spherical overdensity at which an object virializes. Although this is
strictly valid only for $\Omega_{m}=1$ cosmologies, when $\Omega_m$
decreases from $1$ to $0.3$, $\delta_c$ only decreases from $1.686$ to
$\sim 1.675$ \citep{Eke96}. We omit this dependence of $\delta_c$ on
$\Omega_m$, $\Omega_{\Lambda}$ and therefore redshift, for the
$\Omega_m=0.37$ $\Lambda$CDM we are trying to fit here. We also
truncate the mass integral at a specified lower mass limit $M_{\rm
lower}$, which we choose as a free parameter.  In these analytic
estimates, clustering is usually treated with simple linear models in
which the long wavelength cluster distribution is amplified over the
underlying mass distribution by a mass-dependent biasing factor,
leading to a power spectrum $ b(M_1) b(M_2)P(k)$ amplified over the
underlying dark matter spectrum $P(k)$. This turns out to be a small
effect for the SZ power spectra.

 To describe the gas profile, we need the pressure profile. We take
this to be that for an isothermal distribution with a baryon density
profile given by a $\beta$ model with $\beta=2/3$.  For the gas
temperature, we adopt the virial theorem relation given in
\citet{Pen98b}: ${M}/{M_{8}}=(T/T_8)^{3/2} $, where $M_8$ is the mean
mass contained in an $8h^{-1}{\rm Mpc}$ sphere and $T_8 = {4.9
\Omega_m^{2/3}\Omega(z)^{0.283} (1+z)}\, {\rm keV}$. $\Omega(z)$ is
the fraction of matter density with respect to the critical density at
redshift $z$. This relation was obtained by comparing the gas
temperature distribution in simulations with the halo mass function
described above. For the electron number density profile we adopt
$n_{e}=n_{e0} \left[1+ {r^{2}}/{r_{\rm core}^{2}}\right]^{-1}$, scaled
by the central density $n_{e0}$ and core radius $r_{\rm core}$. Since
our simulations show that the pressure profile falls off more rapidly
than the product of temperature and density given here would indicate,
we truncate the pressure at a fraction $fr_{\rm vir}$ of the
`virialized radius' $r_{\rm vir}$. The virial radius is defined as
the radius of a sphere with mass $M$ and mean density $\Delta_c(z)
\bar{\rho}(z)$, where $\bar{\rho}(z)$ is the mean matter density at
redshift $z$ and $\Delta_c$ is given by \citet{Eke96}. If we assume
that gas accounts for a fraction $\Omega_B/\Omega_m$ of the halo mass,
the baryon content fixes one of the 3 parameters, $n_{e0}$. We treat
$f$ and $r_{\rm vir}/r_{\rm core}$ as free parameters to be fit to the
SZ power spectrum.

The bottom panel of Fig.~\ref{fig:szspectra} shows typical fits we can
obtain with the above model using $r_{\rm vir}/r_{\rm core}=4.8$ and
$f=0.9$. Given the number of free parameters, many combinations of
values can yield reasonable fits. In order to investigate the
resolution effects in our simulations we fix $r_{\rm vir}/r_c$ and
$f$ and vary the lower mass cutoff $M_{\rm lower}$ and the $k$
range cutoff in the analytic model. These parameters are related to
the resolution limitations of our simulations. For the \MMH\
simulation, we find the effect of the $k$ range cutoff is negligible,
but that $M_{\rm lower}$ has a larger effect. Since we need more than
100 gas and dark matter particles to resolve a halo, we choose $M_{\rm
lower}$ as the mass of 100 gas particles and 100 dark matter
particles. This corresponds to $0.0015 M_8$ for the $512^3$
simulation. We see substantial deviations developing when $M_{\rm
lower} > 0.01 M_8$. There are as well other uncertainties not included
in our hydro simulations, which are adiabatic. For example, the
outflow of gas from groups could lead to an effective mass cutoff that
is physical rather than resolution dependent. The concentration
parameter for gas, $r_{\rm vir}/r_{\rm core}$ could be a function of mass and
of redshift, which would also change the results.

\section{Simulating CBI Observations of the Sunyaev-Zeldovich
Effect}\label{sec:spectrum}

The estimation of bandpowers used in the analysis of the \CBI\
observation is based on a maximum likelihood technique which assumes
the signal to be a Gaussian random field (Paper~IV). This assumption
appears to be justified in the case of primary CMB anisotropies where
no clear evidence of non-Gaussianity has been found. However, the
contributions from various foregrounds are non-Gaussian. Although the
observations are noise dominated at the scales of interest for the
\CBI\ deep field excess and therefore expected to be predominantly
Gaussian, it is important to test the effect of a small non-Gaussian
signal such as a SZ foreground on the bandpower estimation procedure.
    
We used the simulated SZ maps to test the effect of a SZ foreground on
the \CBI\ analysis pipeline. To do this we constructed detailed mock
data sets using the \CBI\ simulation tools (Paper~IV). We took real
observations for particular dish configurations and pointings and
replaced the observed visibilities by realizations of the expected
signal and instrumental noise. This resulted in simulated observations
with $uv$ coverage identical with the actual observations. The
simulations can also include foreground templates as a distribution of 
point sources or maps of the SZE.

Each simulated $2\times2$ degree SZ map is used as a foreground and
added to a primary CMB background to generate mock observations of the
$08^{\rm h}$ deep field. One point to note is that the \CBI\
observations are actually differenced to minimize any ground pick--up
(Paper II). This involves subtracting the signal from two fields ({\sl
lead} and {\sl trail}) separated by 8 minutes of right ascension on
the sky. In adding a SZ foreground to the simulated observations two
separate SZ maps were used as {\sl lead} and {\sl trail} fields. For
this exercise, we created 20 simulated observations using the 40 \MMH\
maps and 10 for the 20 \SPH\ maps.

The mock data sets were then processed through our power spectrum
estimation pipeline described in Paper~IV. The maximum likelihood
calculation of the power spectrum used template correlations to
project out the expected contribution from point source
foregrounds. Since point source templates were projected out of the
data or were constrained at a known amplitude (as in the case of the
residual unresolved background), the power in the excess was assigned to
the CMB when fitting for the bandpowers since no other correlations
were included in the quadratic power spectrum estimator (except for
instrumental noise).  We reproduced this situation by simulating
observations of primary CMB realizations with SZ foregrounds, but only
allowing for noise and CMB contributions to the total correlation
matrix in the likelihood
\begin{equation}
        {\bf C} = {\bf C}^{\rm N}+{\bf C}^{\rm CMB},
\end{equation}
where the correlations due to the CMB are expressed as a function of
the bandpowers $C_B$ with
\begin{equation}
        {\bf C}^{\rm CMB} = \sum_B {\cal C}_B \frac{\partial {\bf
C}^{\rm CMB}}{\partial {\cal C}_B} \, .
\end{equation}

The SZ maps were produced from simulations of
different cosmologies and we have shown how the spectra can be scaled
to fiducial $\Omega_bh$ and $\sigma_8$ values for comparison. However
when simulating the measurement of bandpowers we chose to use
unscaled contributions in order to test the different amplitude
regimes given by the different values of baryon density. As a primary
CMB signal we used random realizations of a single $\Lambda$CDM model
($\Omega_m=0.30$, $\Omega_b= 0.04$, $\Omega_{\Lambda}=0.70$, $h=0.68$
and $n_s=0.975$).  For the SZ contributions we restricted ourselves to
the $\sigma_8=1.0$ simulations.

\begin{figure}
\plotone{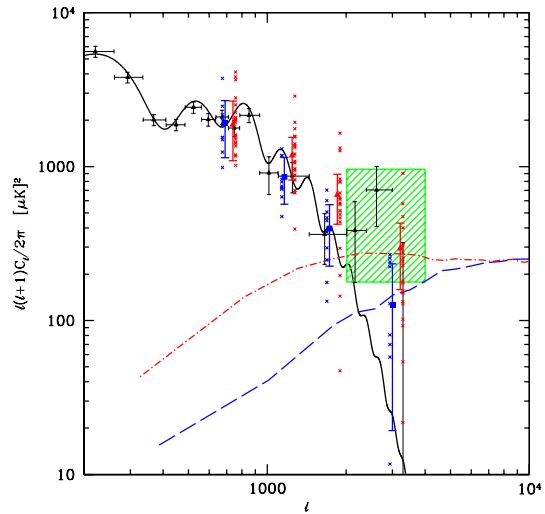}
\caption{Simulated observations of CMB {\sl plus} SZ fields. The
result of bandpower estimation on simulated observations of the $08^{\rm h}$
deep field containing noise, realizations of a fiducial $\Lambda$CDM
model shown in black (solid line) and SZ foregrounds from the 
\SPH\ and \MMH\ $\sigma_8=1.0$ simulations. The blue (dashed)
curve shows the average power spectrum of the input \SPH\ maps and
red (dash-dotted) is the average power spectrum of the input \MMH\
maps. The red (triangle) points are the average bandpowers obtained
from the analysis of 20 separate observations using the \MMH\
maps. The blue (square) points are the average bandpower obtained
from 10 \SPH\ maps. The bandpower estimation pipeline recovers
the correct power in the SZ dominated region $\ell>2000$. The green
(shaded) rectangle shows the 95\% confidence region for the
high-$\ell$ \CBI\ deep field result. Both the \MMH\ and \SPH\
codes show power consistent with the confidence region.}
\label{fig:spectra}
\end{figure}

Results are shown in Fig.~\ref{fig:spectra}. The bandpowers are
averaged over the 20 and 10 independent realizations from the two
$\sigma_8=1.0$ simulations. The triangles (red) are for the
\MMH\ simulations and the \SPH\
results are shown as squares (blue) points. The model used for the
primary CMB contribution is also shown together with the $2\sigma$
confidence region for the deep excess. The errors shown are obtained
from the variance of the measured bandpowers. The averages appear to
recover the input power accurately in both regimes where either the
primary CMB signal or SZ signal dominate. 

The interpretation of the errors is of course complicated in the
SZ-dominated regime by the non-Gaussianity of the SZ signal
(eg. \citealt{Zhang02}). The small area considered in the simulations
results in significant sample variance effects in the measurements of
the SZ power. The scatter in the observed bandpowers over the
different realizations is also shown in Fig.~\ref{fig:spectra}. The
non-Gaussian scatter is significant, although this effect would
decrease for the larger area probed in the joint three-field analysis
and because the high-$\ell$ errors are dominated by the noise. It is
also important to note that if the SZE is a significant source of the
power in the observations, the sample variance component of the errors
derived in the optimal spectrum estimation would be biased.

It is also interesting to note the effect of the low--$\ell$
contribution in the \MMH\ simulations. There is a sizable contribution
to the bandpowers below $\ell\sim 2000$. Although this effect may be
due to the presence of one rather large cluster in the simulation, as
discussed above, it raises the question of whether a SZ signal with
high enough amplitude to explain the excess may already be constrained
by its low--$\ell$ contributions.
                      
\section{Wiener-filtering the \CBI\ Deep Field Observations}\label{sec:maps}

The \CBI\ does not have sufficient frequency coverage to strongly
distinguish different signals by their frequency dependence (e.g.,
CMB, SZ, and Galactic foregrounds). Spectral separation of the
different components is therefore infeasible. Although the sensitivity
of the CBI results presented here is not sufficient to conclusively
identify individual features in the maps, we present examples and
simulations of a Wiener-filtering technique described in Paper~IV. 

The total mode to mode correlations in the data are separated into four
components in a typical analysis of the deep field visibilities with
the weight matrix ${\bf W}$ as the sum
\begin{equation}
{\bf W} = ({\bf C}^{\rm N}+{\bf C}^{\rm CMB}+{\bf C}^{\rm SRC}+{\bf C}^{\rm res})^{-1},
\end{equation}
with noise (N), CMB, known source (SRC) and residual source (res)
components, respectively. (In papers II and III, the known source
contribution is split into two components, for sources with and
without measured flux densities.)

In the limit where all the components can be considered Gaussian
random fields we can define the probability of each signal given the
data as the Gaussian \citep{Bond02}
\begin{eqnarray}
\ln P(\Delta^{\rm X}|\bar\Delta) &=& - \frac{1}{2} (\Delta^{\rm X}-\langle
\Delta^{\rm X}| \bar\Delta \rangle)^{\dagger}({\bf C}^{\rm X})^{-1}(\Delta^{\rm X}-\langle
\Delta^{\rm X}| \bar\Delta \rangle)\nonumber\\
& & -\frac{1}{2}{\rm Tr}\ln {\bf C}^{\rm X}-N\ln\sqrt{2\pi}, 
\end{eqnarray}
where $\bar\Delta$ are the observations, $\Delta^{\rm X}$ is the 
map vector of the component ${\rm X}$, ${\bf C}^{\rm X}$ is the mode to
mode correlation matrix for the component and $N$ is the number of
modes in the map. The mean $\langle
\Delta^{\rm X}| \bar\Delta \rangle$ defines the Wiener-filtered map
of the component given the observations
\begin{equation}
        \Delta^{\rm X} = {\bf C}^{\rm X}{\bf W}\bar\Delta,
\end{equation}
with variance about the mean
\begin{equation}
        \langle\delta\Delta^{\rm X}\delta\Delta^{{\rm
            X}\dagger}|\bar\Delta\rangle 
= {\bf C}^{\rm X}-{\bf C}^{\rm X}{\bf W}{\bf C}^{\rm X}.\label{eq:variance}
\end{equation}

In our case, the maps $\Delta$ are column vectors containing the
gridded visibility estimators (Paper~IV), and sky plane maps can
be obtained by Fast Fourier Transforming these estimators in the
$uv$--plane. The correlation structure of the estimator grids are a
necessary by-product of our gridding method and power spectrum
estimation pipeline. The calculation of the correlations includes all
the relevant information on the sampling structure and convolution of
the fine-grained observation plane. Eq.~\ref{eq:variance} therefore
includes all aspects of the uncertainties in the Wiener-filtered map
resulting from the observations.

\begin{figure}
\begin{center}
\plotone{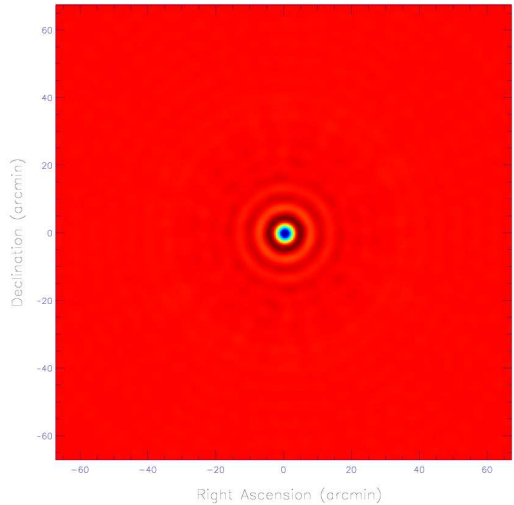}
\end{center}
\caption{The image produced by gridding a simulated observation of a
point source placed at the center of the $08^{\rm h}$ field.}
\label{fig:psf}
\end{figure}

Due to the complex weighting applied to the observed visibilities in
our gridding scheme it is difficult to assign units to the maps
produced by simply Fourier transforming the gridded {\it
uv}--plane. In order to obtain images with approximate normalizations
in mJy/pixels, we reweight by dividing the vector at each lattice site
by a second vector obtained by gridding a 1 mJy point source placed at
the center of each field. It is important to note that this
deconvolution is carried out in the coarse grained lattice and is not
a full deconvolution of the effect of the {\it dirty beam} since no
extra information is added to complete the {\it uv}--plane when
gridding. In Fig.~\ref{fig:psf} we show the image obtained by Fourier
transforming the gridded {\it uv}--plane of a simulated observation of
a point source placed at the center of the $08^{\rm h}$ field.

\begin{figure*}
\begin{center}
\begin{tabular}{cc}
\makebox[3in][c]{\includegraphics[width=2.5in]{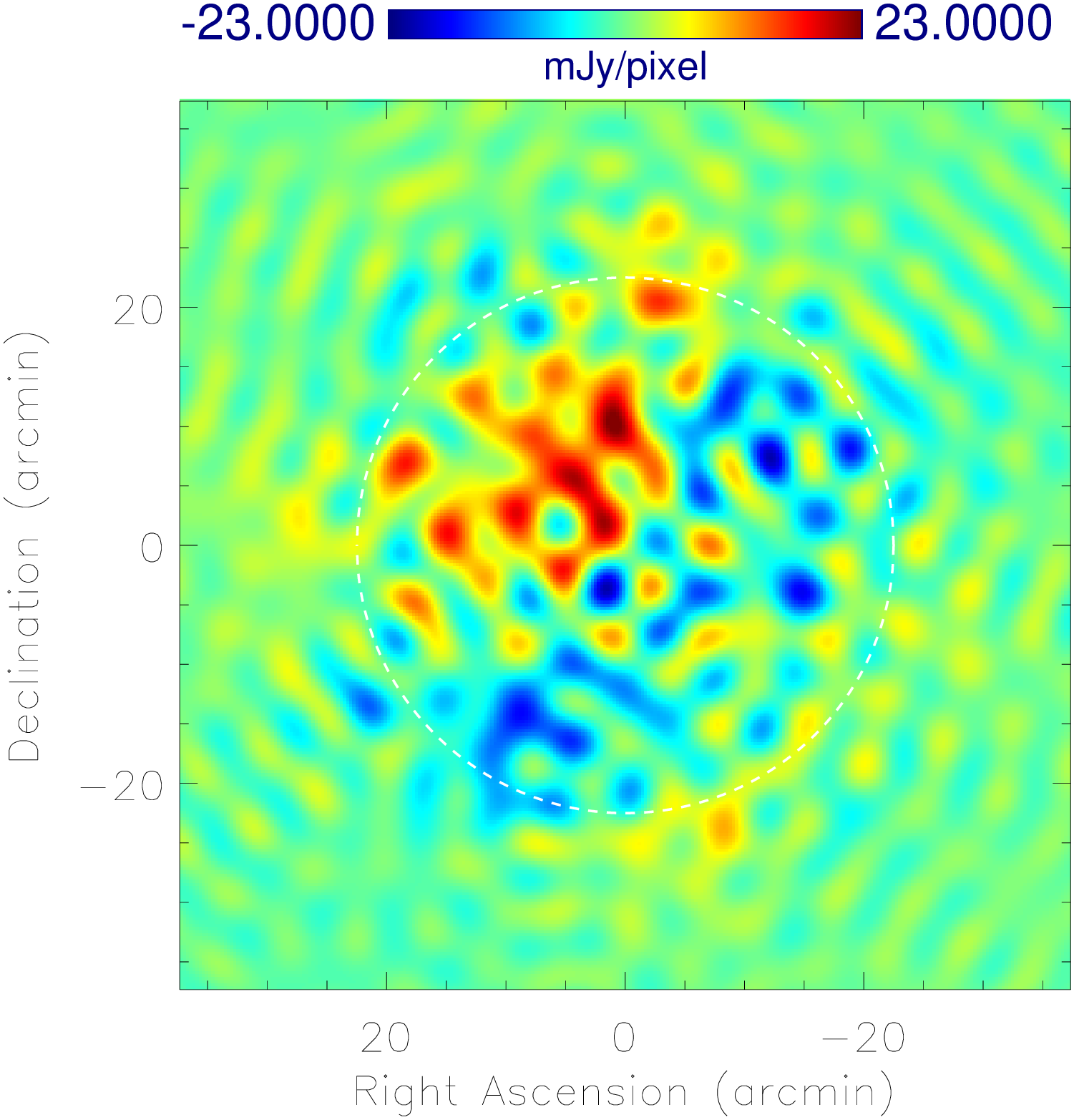}} &
\makebox[3in][c]{\includegraphics[width=2.5in]{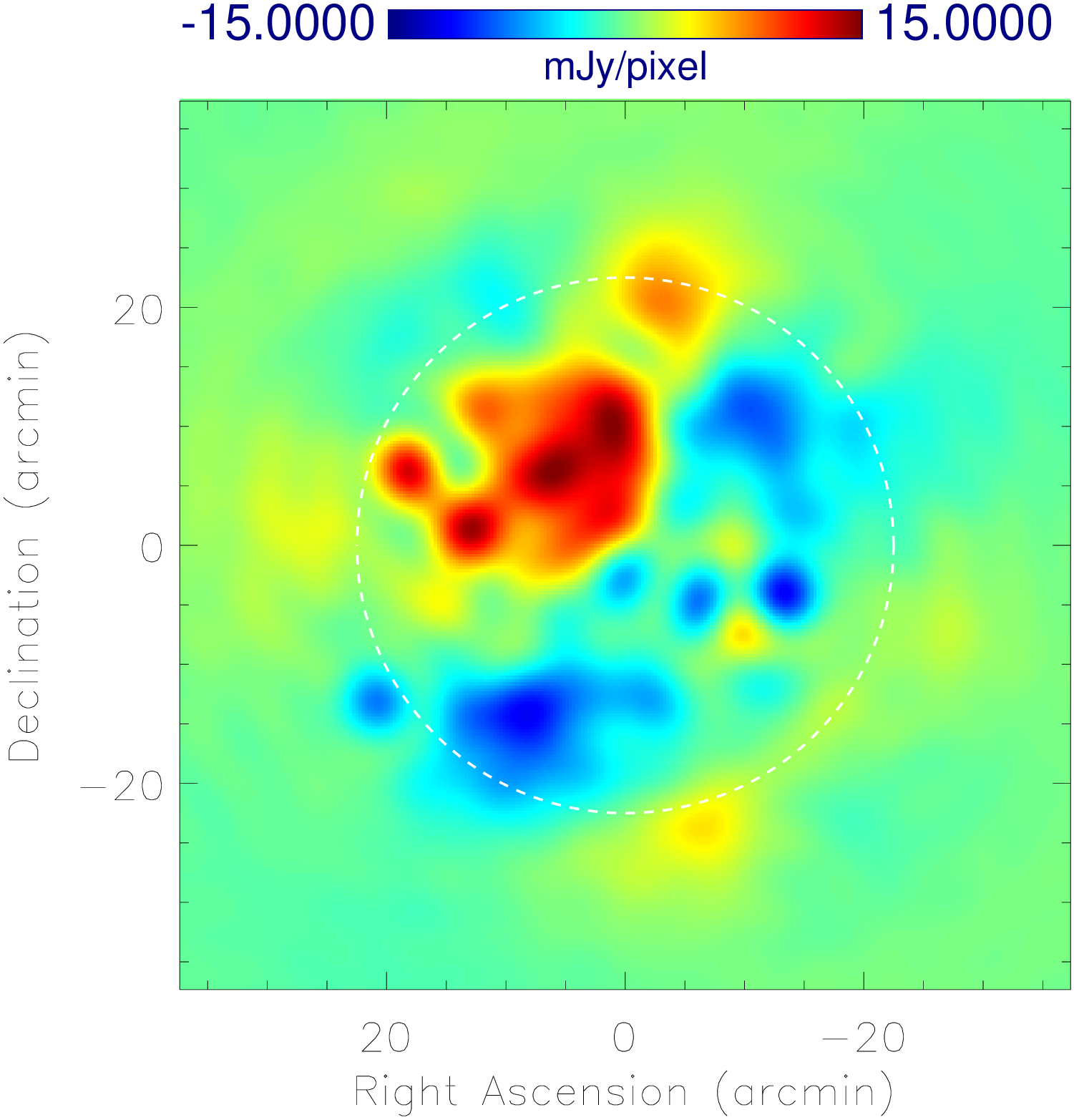}} \\[1cm]
\makebox[3in][c]{\includegraphics[width=2.5in]{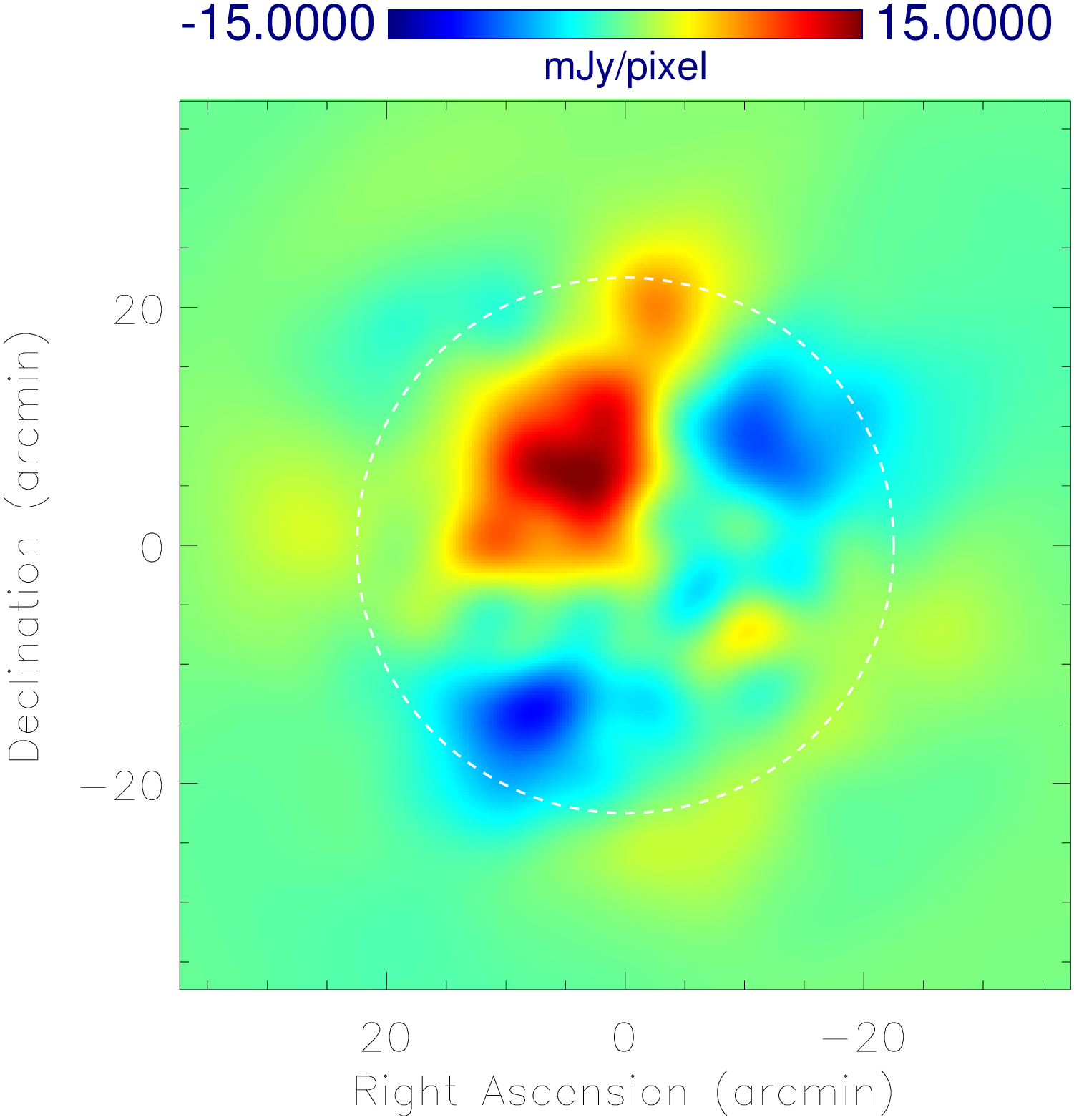}} &
\makebox[3in][c]{\includegraphics[width=2.5in]{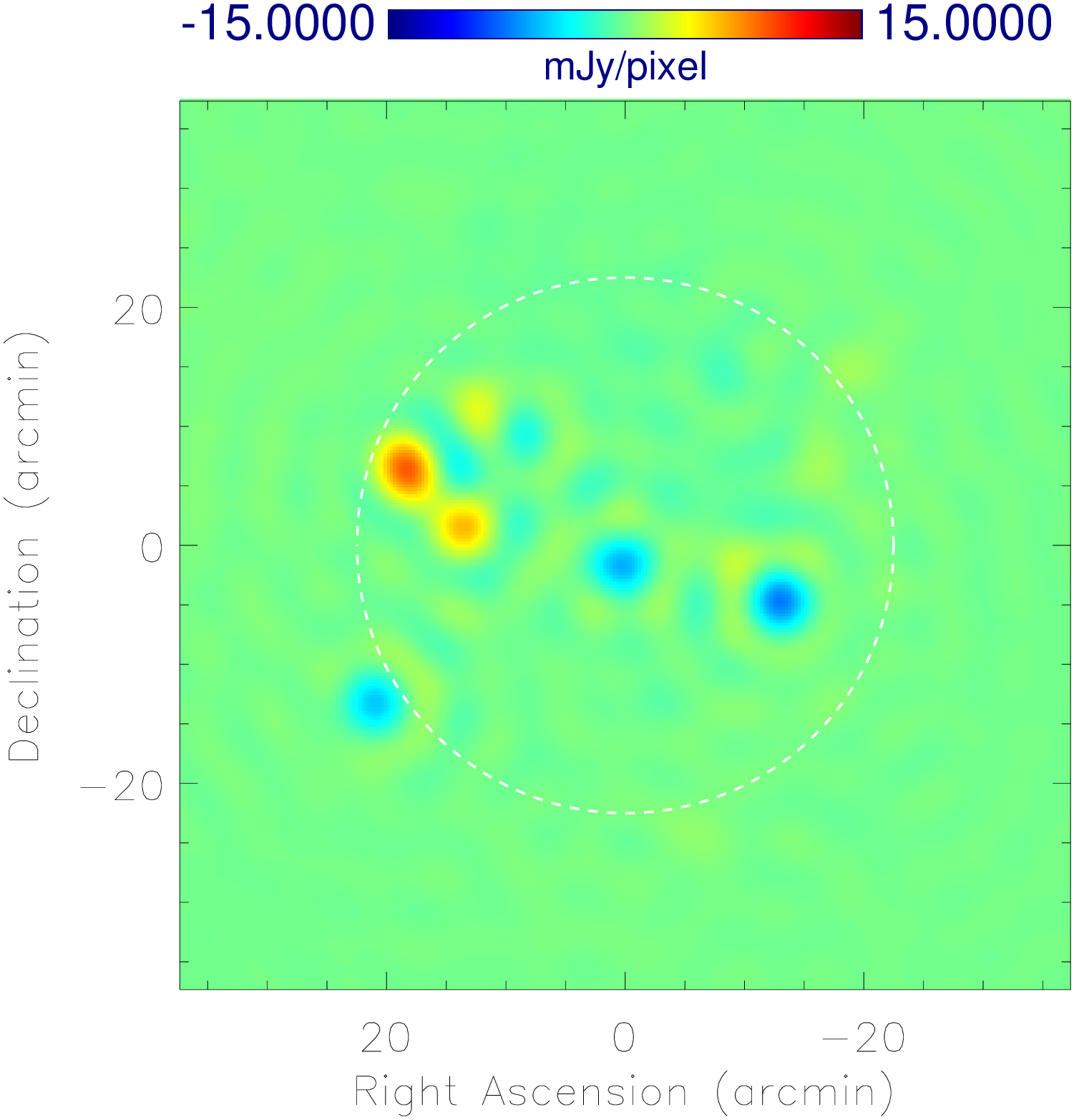}} \\[1cm]
\end{tabular}
\end{center}
\caption{Filtered images of the \CBI\ $08^{\rm h}$ deep field. The
total signal {\sl plus} noise image (top left) displayed as a
$70^\prime$ by $70^\prime$ field. The white (dashed) circle shows the
$45^\prime$ FWHM of the primary beam. A total signal image (top right)
is obtained by using the sum of CMB and point source foreground
correlations as a template. The amplitude of the residual source
background is set to a flux of $0.08 {\rm Jy}^2{\rm sr}^{-1}$ in the
filter. Using the joint deep field bandpowers as amplitudes in the
filters we obtain the optimal image for the CMB (bottom left). This
image encompasses the power attributed to the high-$\ell$ excess. The
known point source residuals can also be filtered out of the data
(bottom right). Colour scales are kept constant for the filtered
images to ease comparison of the relative amplitudes of the
components. The images are deconvolved by the response of a simulated
source placed at the center of the field.}
\label{fig:wiener}
\end{figure*}

We show an example of the use of such filters in
Fig.~\ref{fig:wiener}. The sequence shows the result of applying
different filters to the $08^{\rm h}$ deep field observation. The
original image (upper left) is first filtered to obtain a map of the
total signal contributions which includes CMB and point sources (upper
right). The CMB power spectrum obtained from the joint deep field
analysis is then used as a template to obtain a map of the CMB
contribution (lower left). The template includes the excess power
above $\ell\sim 2000$. The post-subtraction residuals in the known
point sources can also be separated into an image (lower right) which
shows the total contribution of the modes which are projected out when
estimating the power spectrum. The mean Wiener-filtered maps make no
statement on the significance of any feature by themselves and
considering only the mean can be misleading since there is no
information on the allowed fluctuations around the mean in the map
itself. This is particularly so in the case of interferometric
observations where the nature of the noise and sampling uncertainties
means that, although the power due to single features is conserved
their structure in the sky plane is complicated by the extended nature
of the correlations in the uncertainties. As shown above, however, in
the Gaussian limit we can assign confidence limits on any features
using Eq.~\ref{eq:variance}.

A useful method of visualizing the significance of the features is by
creating constrained realizations of the fluctuations and comparing
these to the mean. For example we can obtain a random realization of
the fluctuations by taking $\Delta_i^{\rm X} = ({\bf C}^{\rm X}-{\bf
C}^{\rm X}{\bf W}{\bf C}^{\rm X})^{1/2} \xi$ where
$\xi$ are random Gaussian variates with unit
norm. Adding these maps to the mean we get an idea of the allowed
fluctuations about the mean. Strongly constrained features will be
relatively unaffected by the fluctuations and represent the high
signal--to--noise ratio limit where ${\bf C}^{\rm X}-{\bf C}^{\rm
X}{\bf W}{\bf C}^{\rm X} \rightarrow 0$. Features measured at levels
comparable to the `generalized' noise of the observations will be
washed out by the fluctuation levels and represent the limit ${\bf
C}^{\rm X}-{\bf C}^{\rm X}{\bf W}{\bf C}^{\rm X} \rightarrow {\bf
C}^{\rm X}$. The full pixel--pixel covariance matrix of the component
map can be calculated by direct FFT of the covariance matrix or by
taking Monte Carlo ensemble averages of the fluctuation realizations.

\begin{figure*}
\epsscale{0.7}
\plotone{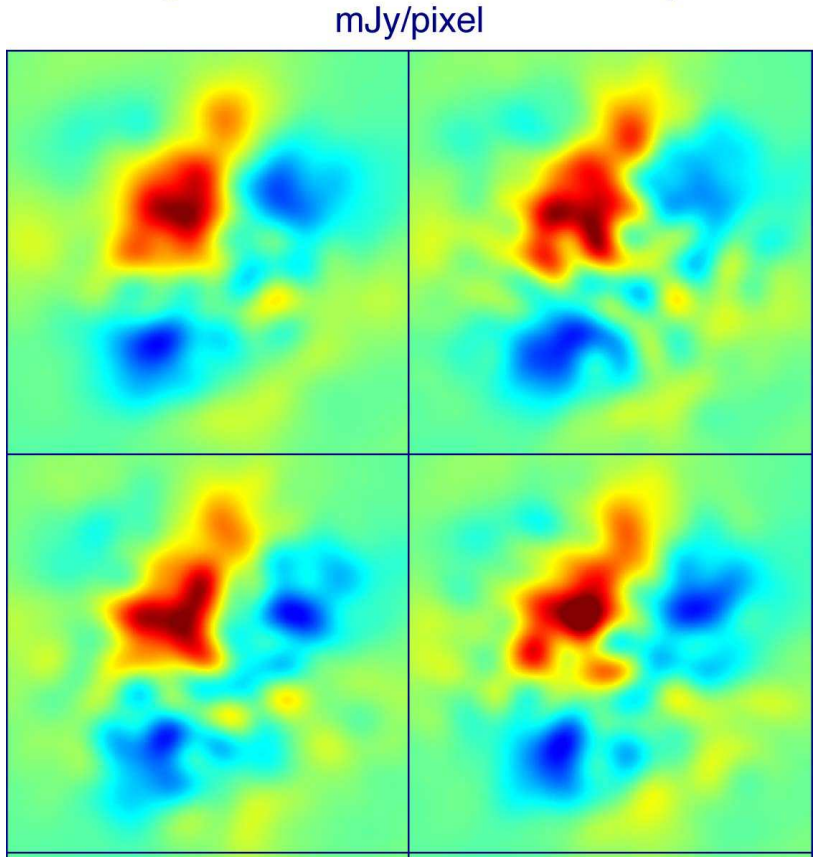}
    \caption{Another way to illustrate whether the features seen in
      the Wiener-filtered images are robust is to show a few
      fluctuations about the mean (Wiener) map, here for the
      $08^{\rm h}$ deep field. The top left panel reproduces the mean
      CMB component map of Fig.~\ref{fig:wiener}. The remaining three
      panels show the addition of three constrained realizations of
      the allowed fluctuations to the mean map. At the
      signal--to--noise ratios of the larger scale CMB observations
      the features are robust to the fluctuations. Note that these
      images are `dirty maps': no attempt has been made to compensate
      for gaps in the $uv$-coverage. The features are a combination of
      main-lobe and side-lobe responses to the CMB. }
\label{fig:fluc}
\epsscale{1.0}
\end{figure*}

The level of fluctuations allowed around the mean for the CMB
component of Fig.~\ref{fig:wiener} is shown in Fig.~\ref{fig:fluc}.
The four panel figure shows the Wiener filter of the component and the
same map with three random realizations of the fluctuations added to
it. As expected the CMB component is well constrained on the scales
shown in the image since the signal--to--noise ratio is high: {\it
i.e.} the main features in the map are stable with respect to the
fluctuations. The features are therefore expected to be closely
related to `real' structures on the sky; but the nature of the
observations, in particular the side-lobes of the point spread
function (synthesized beam), means that the detailed structure of the
features is not well constrained and most likely does not reflect the
precise shape of the objects. The filtering of the large scale power
by the interferometric observations also makes the correspondence with
the large scale features on the sky not as intuitive as with sky plane
based measurements.

Using the specific SZ models we introduced in \S~\ref{sec:spectrum} we
can now extend the image analysis by template-filtering the images
specifically for SZ contributions. As described above the only
components modeled in the total correlation matrix are the CMB,
instrumental noise and source components. We therefore construct the
SZ template as
\begin{equation}
        {\bf C}^{\rm SZ} = \sum_B {\cal C}^{\rm SZ}_B \frac{\partial {\bf
C}^{\rm CMB}}{\partial {\cal C}_B}.
\end{equation}
where the amplitudes ${\cal C}^{\rm SZ}_B$ are obtained by filtering the
average SZ power spectra by the deep field band window functions (see
 Paper~II).

\begin{figure*}
\begin{tabular}{ccc}
\makebox[2.0in][c]{\includegraphics[width=1.8in]{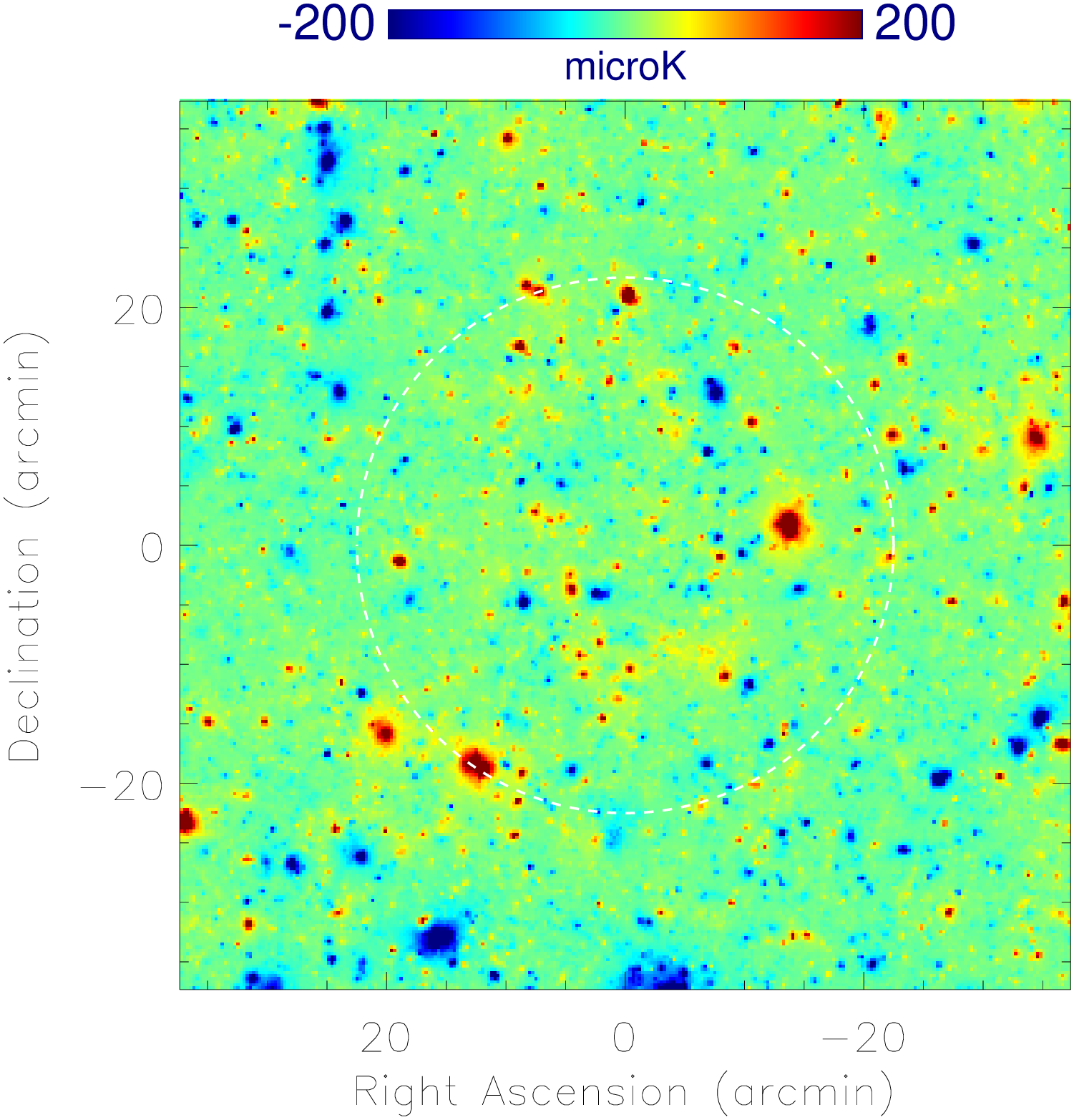}} &
\makebox[2.0in][c]{\includegraphics[width=1.8in]{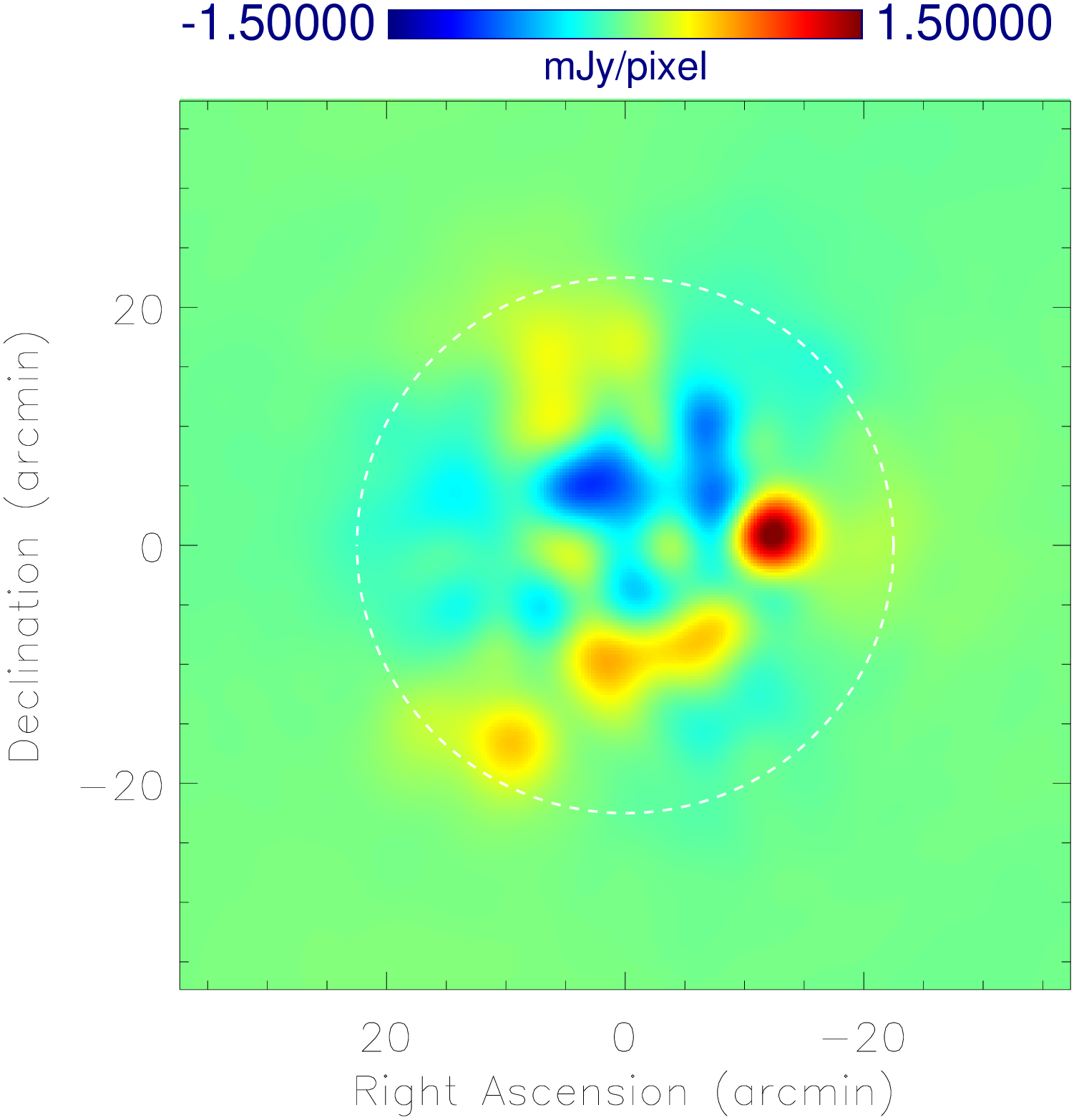}} &
\makebox[2.0in][c]{\includegraphics[width=1.8in]{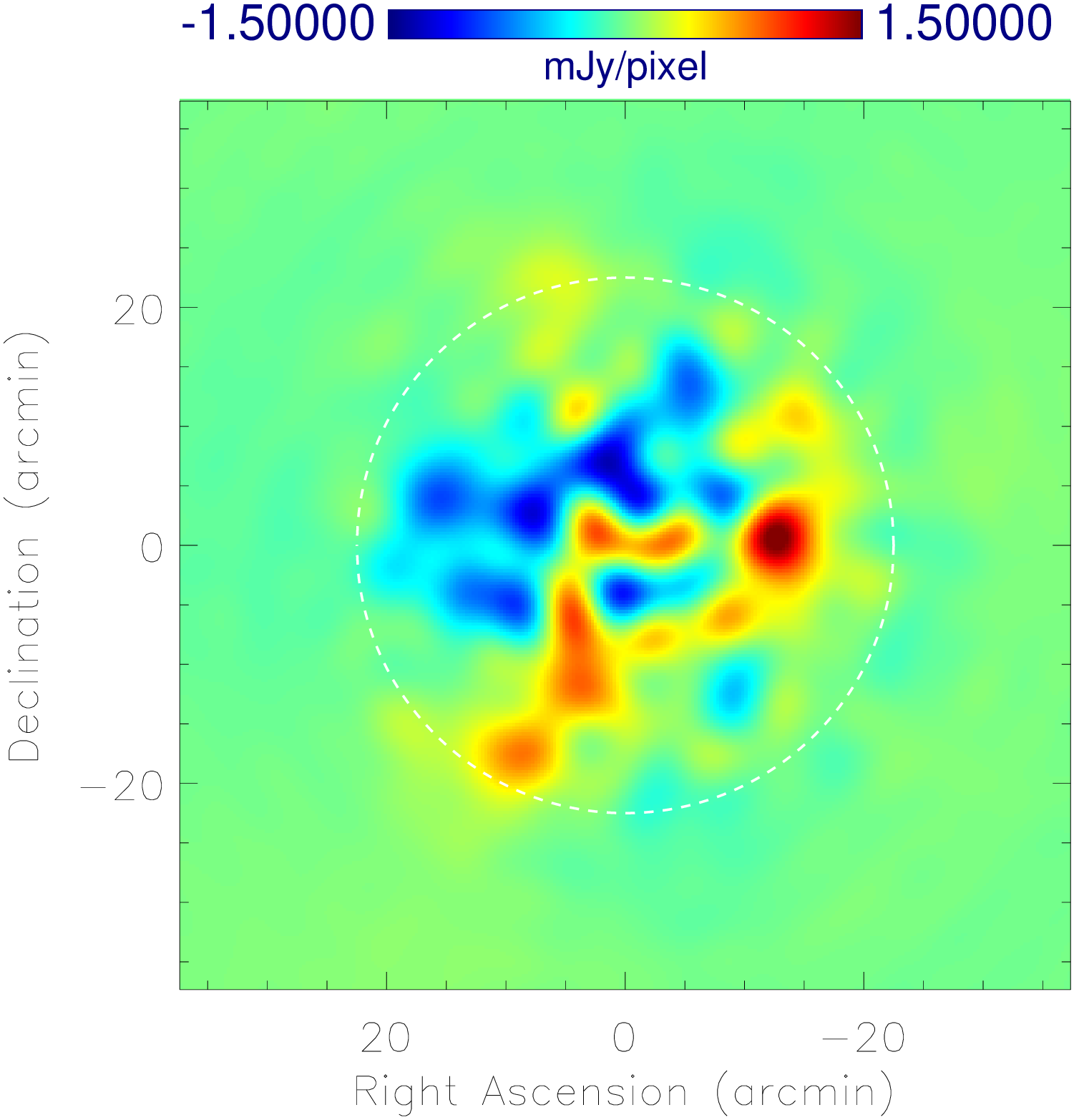}}\\[1cm]
\makebox[2.0in][c]{\includegraphics[width=1.8in]{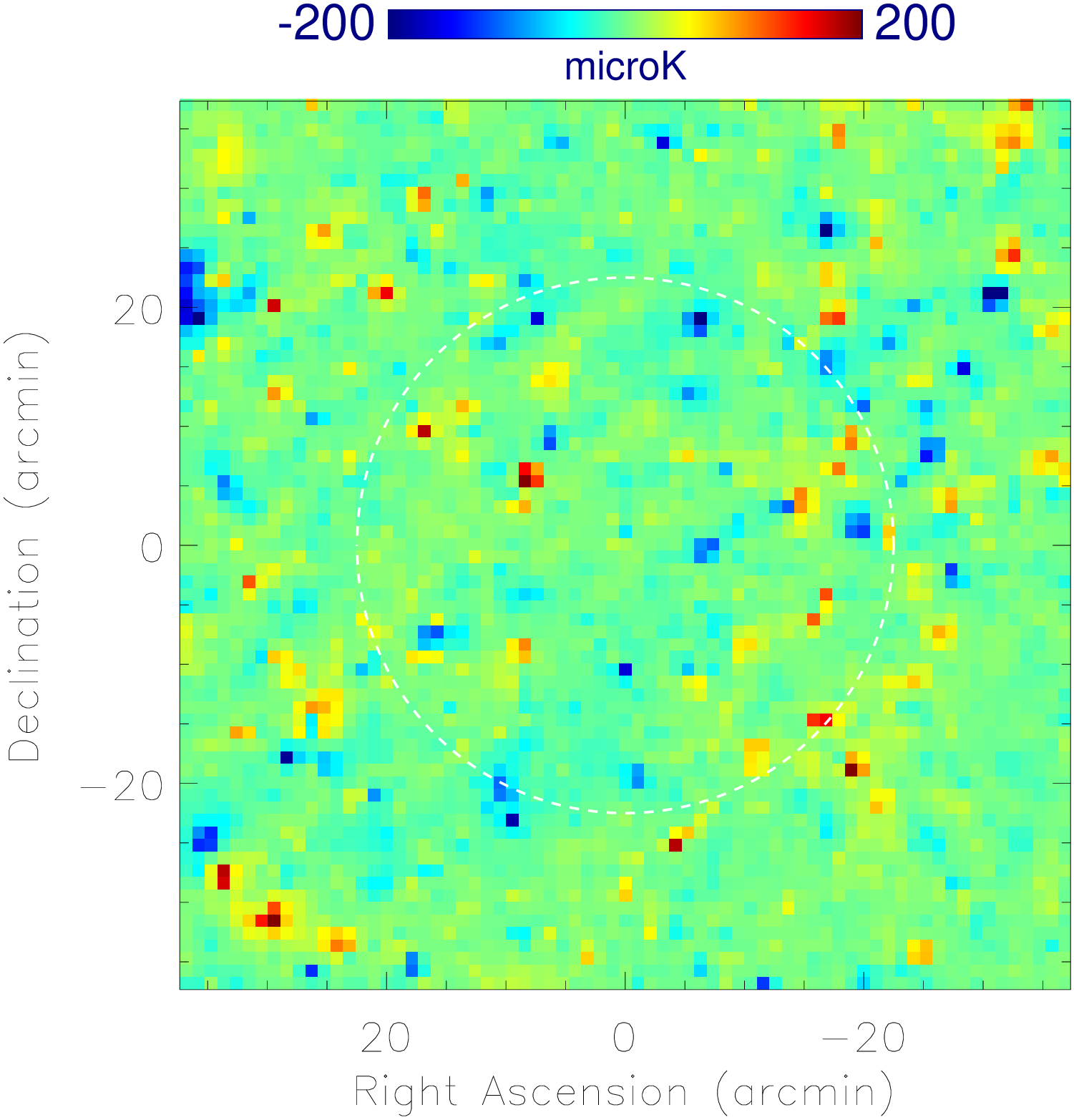}} &
\makebox[2.0in][c]{\includegraphics[width=1.8in]{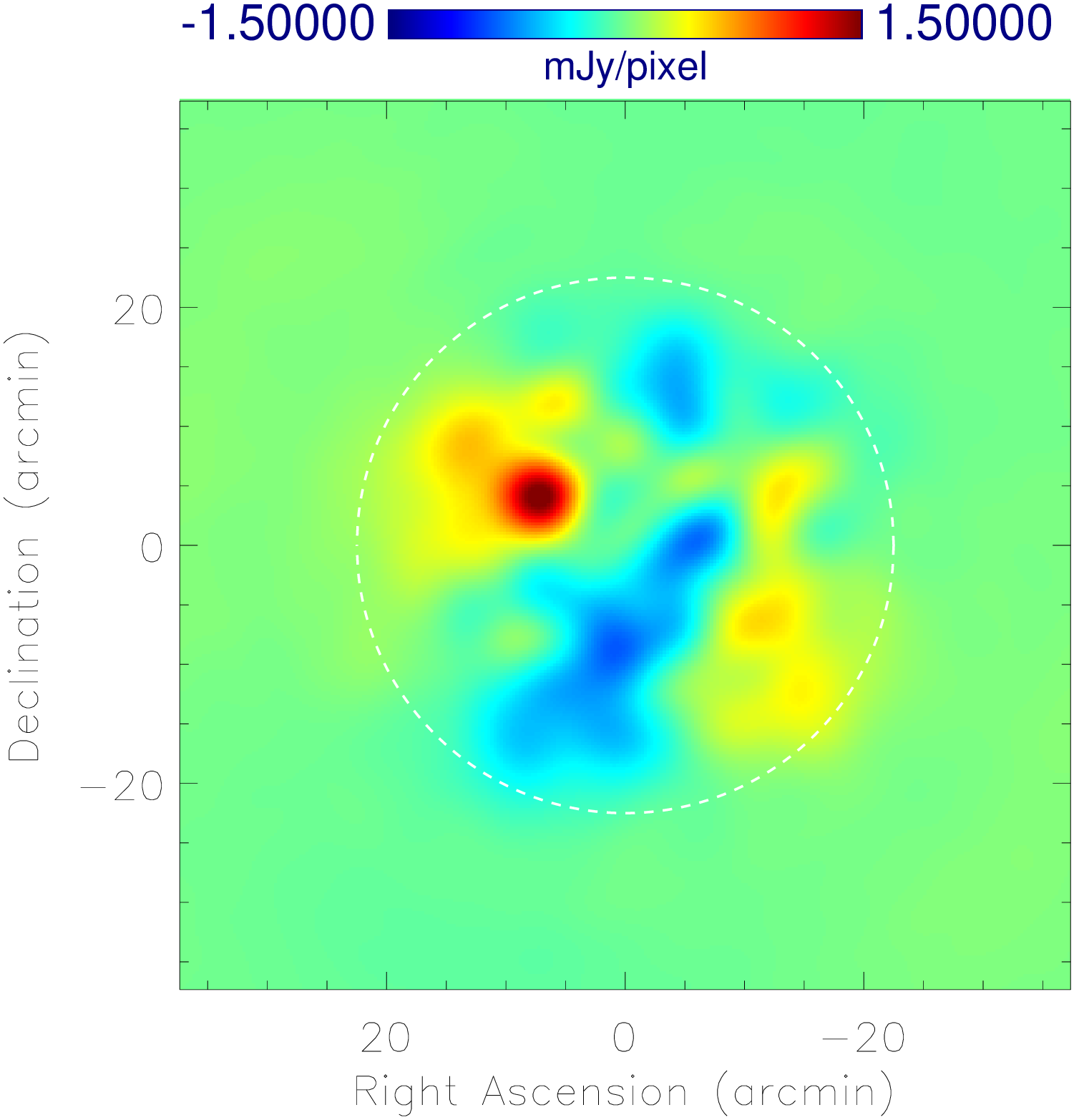}} &
\makebox[2.0in][c]{\includegraphics[width=1.8in]{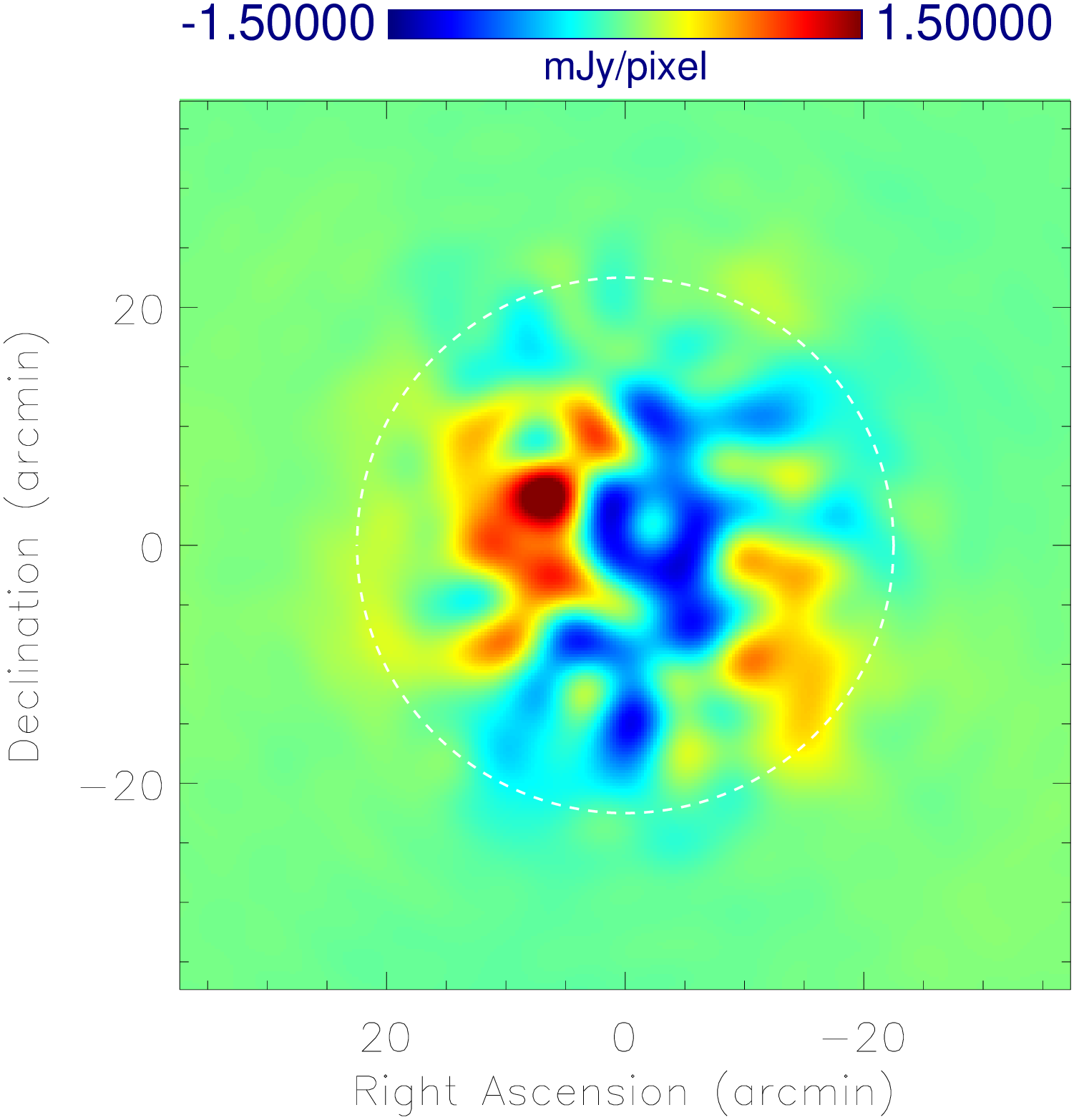}} \\[1cm]
\end{tabular}
\caption{Simulated observations of SZ signals. A single SZ realization
is shown for the \SPH\ (top left) and \MMH\ (bottom left)
simulations. The area represents a ({\sl lead-trail}) differenced
field used as the signal in a simulated observation of the $08^{\rm
h}$ deep field. The maps are used to generate mock data sets with
exact replications of the $uv$--coverage as the real data which are
then filtered using the respective SZ templates. Mocks with close to
negligible noise levels (top and bottom center) show the direct
representation of the \CBI\ processed input maps. The same realization
but with the same noise levels as the observed data is in the top and
bottom right panels. To within the accuracy allowed by the noise, the
filtered images successfully reproduce the features of the input maps
within the primary beam area.}
\label{fig:szinout1}
\end{figure*}

The resulting SZ-filtered images should reproduce the structure in
input SZ maps and we find that this is indeed the case. The sequence
of images in Fig.~\ref{fig:szinout1} is an example of the filtering
for one of the \SPH\ maps (top row) and one of the \MMH\ maps (bottom
row). The left-hand panels show the input SZ maps ({\sl lead}$-${\sl
trail}) and the middle ones show the result of filtering mock data sets
where the noise has been reduced to negligible levels. These
provide a control set to compare with the
reconstructed noisy observations shown on the right. The strongest
features in both images reproduce the strongest SZ signatures in the
input templates, although one has to take into account the tapering
effect induced by the primary beam when comparing the input maps to
the filtered data (see Paper~IV). The tapering suppresses the
amplitude of significant features further out in the beam with respect
to those close to the center of the field. The noise level is such
that the significance of the `detections' is not obvious. These can be
quantified by comparing with the allowed fluctuations about the mean
Wiener filter (Eq.~\ref{eq:variance}).

\begin{figure*}
\begin{center}
\begin{tabular}{cc}
\makebox[3.0in][c]{\includegraphics[width=2.5in]{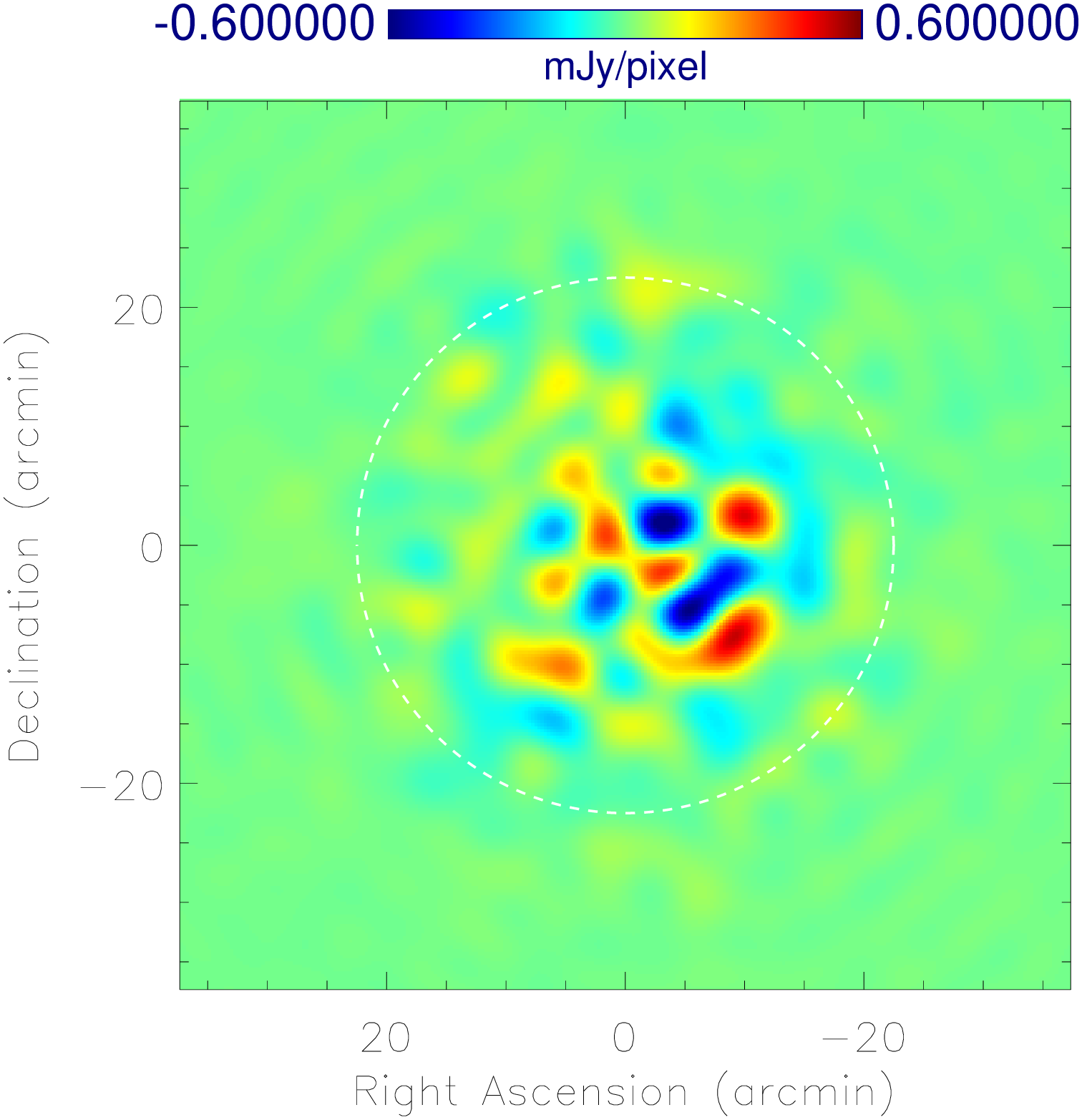}} &
\makebox[3.0in][c]{\includegraphics[width=2.5in]{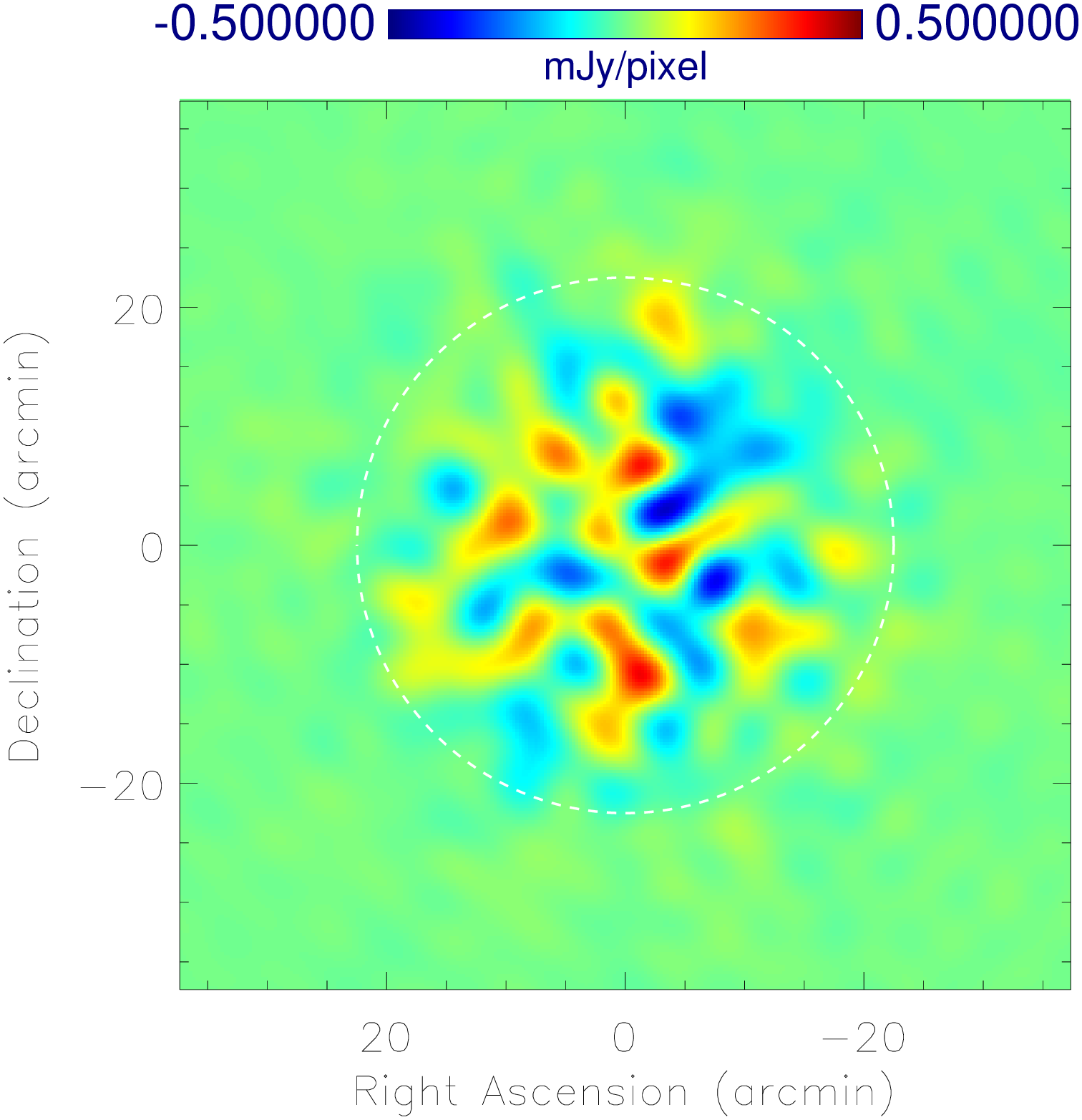}}\\[1cm]
\makebox[3.0in][c]{\includegraphics[width=2.5in]{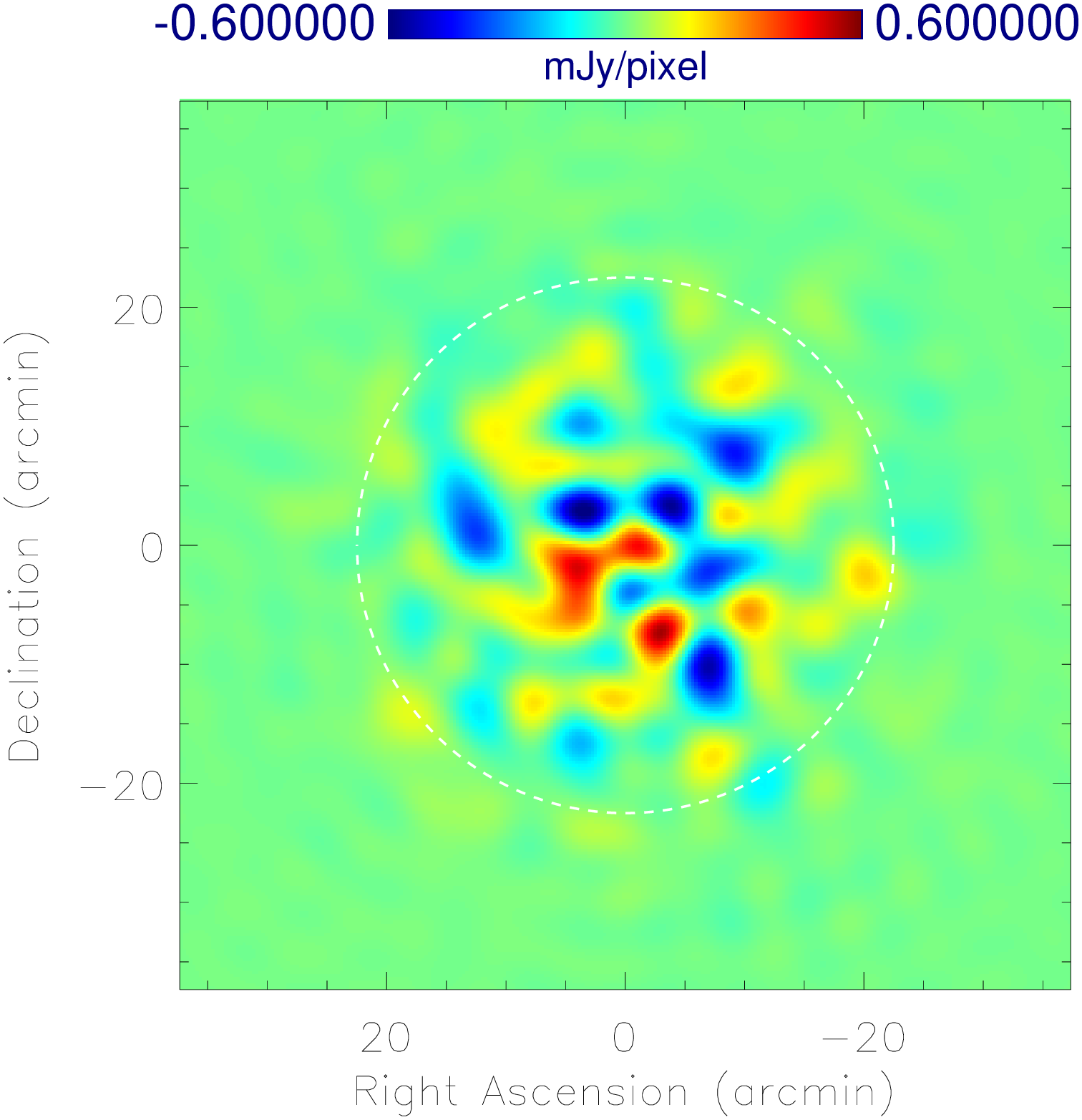}} &
\makebox[3.0in][c]{\includegraphics[width=2.5in]{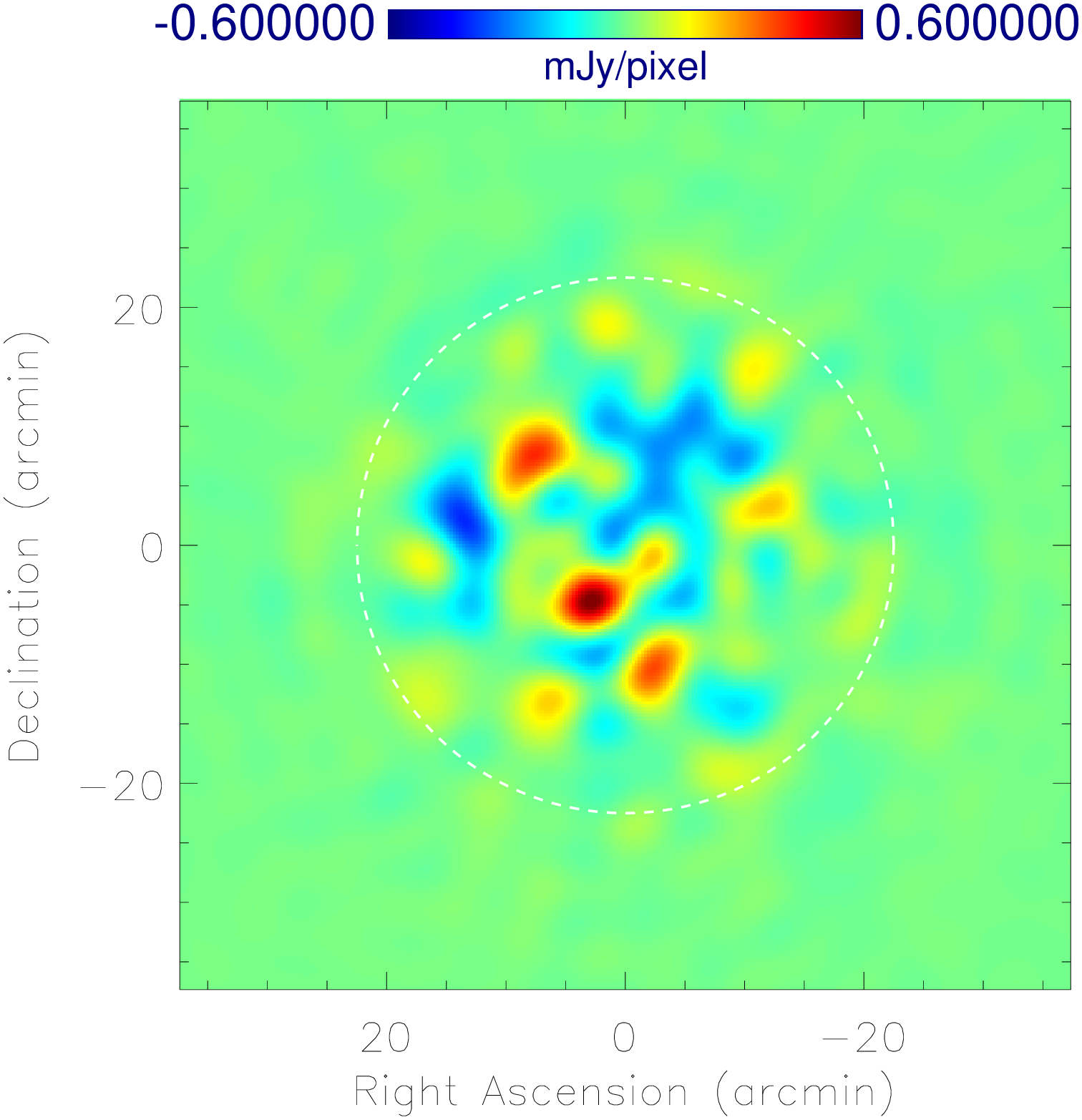}}\\[1cm]
\end{tabular}
\end{center}
\caption{The split deep data Wiener-filtered with the SZ template, for
the $08^{\rm h}$ field (top) and the $20^{\rm h}$ field
(bottom). There is no clear evidence of a high signal--to--noise
stationary feature.}
\label{fig:szhalves}
\end{figure*}

We have also applied the SZ Wiener filter to the deep field data
itself. These SZ-filtered images scale linearly with the overall
normalization of the SZ power spectrum. Hence, by calibrating the SZ
template filters with the observed deep high--$\ell$ power, we can
obtain an image with the same signal--to--noise ratio as the
observations. In fig~\ref{fig:szhalves}, we show the effect of the SZ
filter on the data split in two roughly equal halves for the $08^{\rm
h}$ and $20^{\rm h}$ deep field observations. When comparing the two
halves the noise will be uncorrelated, so any commmon stationary
signal of sufficient amplitude should appear in both halves. One sees
no obvious strong signal-dominated stationary features, but there is
one in each field, albeit with amplitudes comparable to the noise
fluctuations. We conclude that the images are noise-dominated and
there are no obvious features that could single--handedly account for
the observed excess. This is in agreement with the excess being a {\it
statistical} measurement as opposed to a high signal--to--noise
observation.

\section{Discussion}\label{sec:disc}

We have presented estimates for the possible contribution from the SZ
effect to the \CBI\ deep field observations which show evidence of an
excess in power over standard primary CMB scenarios above $\ell\sim
2000$. Our numerical simulations show that an amplitude of
$\sigma_8\approx 1$ appears to give, on average, enough power in the
SZ signature to explain the excess. Current primary CMB and LSS data
favor models with somewhat lower normalizations: for flat, HST--$h$
and LSS priors, $\sigma_8=0.88^{+0.08}_{-0.08}$ for the June 2002
dataset and $\sigma_8=0.86^{+0.04}_{-0.03}$ for the March 2003
compilation. Recent weak-lensing-only results, {\it e.g.},  $0.83 \pm 0.07$
\citep{Ludo05}, are also quite compatible with CMB-only values.

In assessing whether $\sigma_8\approx 0.9$ is too low to explain the
anomaly, the non-Gaussian nature of the sample variance of the SZE
should be included. This issue has been partially addressed in the
literature
\citep{CooraySheth02,Springel01,Komatsu:2002wc}. \citet{Goldstein02}
have shown that a joint analysis of our \CBI\ deep data with
recent data from ACBAR \citep{Kuo02} and BIMA \citep{BIMA02} which
simultaneously fits the amplitudes of the primary and SZ power spectra
and roughly accounts for the non-Gaussianity of the SZE requires
$\sigma_8 = 0.96^{+0.08}_{-0.16}$, with a maximum likelihood at 1.00,
consistent with the constraints on $\sigma_8$ from primary CMB and LSS
data, as illustrated in Figures~\ref{fig:priors8} and
\ref{fig:priors8Om56a}.  \citet{Readhead04} applied a similar joint
analysis to that in \citet{Goldstein02} to the bandpowers of the
ACBAR, BIMA, and CBI 2-year data. An attempt to bracket the effects of
non-Gaussianity was made by allowing the sample-variance component of
the error bar to increase by a factor of up to four: the estimate of
$\sigma_{8}$ hardly changed, going from $0.98^{+0.06}_{-0.07}$ to
$0.98^{+0.06}_{-0.08}$. \citet{Goldstein02} used a factor of three to
estimate the non-Gaussian effect.  A bigger impact is found at the
$2\sigma$ level: with a uniform prior in $\sigma_8$ there is a tail
skewed to low $\sigma_8$, primarily because the amplitude being fit by
the analysis is $\sigma_8^7$, and the distribution is roughly Gaussian
in that variable, as shown explicitly in Fig.~12 of
\citet{Readhead04}. We note that the errors are predominantly from
noise and not sample-variance for the CBI data, which is why variation
by a factor of four does not have a huge effect.

The \citet{Goldstein02} and \citet{Readhead04} analyses took into
account the competing effects of the damping tail of the primary CMB
fluctuations and the rising SZ power spectrum over the entire $\ell$
range, and not just in the $\ell > 2000$ regime where the ``excess''
is clearly seen. The assumption is that the power spectra are
described by an offset-log-normal distribution, which has been shown
to hold well for Gaussian signals and noise \citep{BJK2000,Sievers02}.
The distribution becomes a Gaussian one in the noise-dominated regime
and a log-normal one in the signal-dominated regime. If the signal
comes from clusters and groups, the latter will not be accurate,
and further work is needed to properly assess the limits derived from
the ``excess''.  What is needed is a full suite of simulations, guided
by semi-analytic theory, to Monte Carlo the effects of non-Gaussianity
rather than rely on expanded signal-errors.

The excellent agreement among the runs using the two separate
hydrodynamic algorithms is encouraging and lends support to the
conclusions on the required normalization. We note that the
simulations employed in this work were developed and run independently
from each other and the \CBI\ analysis effort
\citep{Bond01,Zhang02}. However, both simulations were adiabatic, with
entropy increases occurring only by shock heating. The influence of
cooling and entropy injection from supernovae, etc., on the predicted
SZ spectrum remains to be explored, as does the effect of such
technical issues as numerical convergence as the simulation resolution
is changed.

The limited number of mock simulations we have carried out show that
our experimental setup does reproduce correctly the signal of
simulated SZ foregrounds. This suggests the Gaussian assumption
implicit in our bandpower estimation procedures does not break down
at the SZ signal-to-noise ratios of the deep field
observations.

If the excess is not due to the SZE, what could it be? Other possible
origins of the excess were addressed in Paper II. It is inconsistent
with adiabatic inflationary predictions for primary anisotropies on
the damping tail only at the $\sim 3\sigma$ level. The excess clearly
needs to be further explored, both within the CBI dataset itself, and
also by correlating with other observations of these deep fields. The
statistical nature of the detected excess means we cannot identify
single features responsible for the excess in the deep fields even
after Wiener-filtering the data with SZ derived templates.

Our principal conclusion is that the \CBI\ excess could be a result of
the SZ effect for the class of $\Lambda$CDM concordance models if
$\sigma_8$ is in the upper range of values allowed by current CMB
data. The simple ${\cal C}_\ell \propto \sigma_8^7 (\Omega_bh)^2$
scaling (eq.~\ref{eq:scaling}) shows that the lower $\sigma_8$ values
preferred by the CMB data imply SZ signals below, but not too far
below, the sensitivity obtained by the \CBI\ deep observations. This
work also highlights how the signature of the SZ effect has great
potential for constraining the amplitude $\sigma_8$. The sensitive
scaling of the CMB power with $\sigma_8$ will result in significant
constraints even with large errors on the observed bandpowers. This
further underlines the significance of `blank field' observations at
$\ell \gta 2000$ which should reveal the SZ structure that necessarily
lurks as a byproduct of CMB-normalized structure formation models.

\acknowledgements We thank Hugh Couchman, Tom Quinn and Joachim Stadel
for interactions on the computational issues. Research in Canada is
supported by NSERC and the Canadian Institute for Advanced
Research. The computational facilities (SHARCNET at McMaster and
PSciNet at Toronto) are funded by the Canadian Fund for Innovation.
This work was supported by the National Science Foundation under
grants AST 94-13935, AST 98-02989, and AST 00-98734. We are grateful
to CONICYT for granting permission to operate the \CBI\ at the
Chajnantor Scientific Preserve in Chile.


\begin{thebibliography}{}

\bibitem[Allen et al.(2003)]{Allen03}Allen, S.W. Schmidt, R.W. Fabian,  A.C.  \& Ebeling,   H. 2003, \mnras\ 342, 287 (astro-ph/0208394)

\bibitem[Atrio-Barandela \& Mucket(1999)]{Atrio99}Atrio-Barandela, F. \& Mucket, J., 1999, \apj, 515, 465

\bibitem[Bacon et al.(2002)]{Bacon02}Bacon, D., Massey, R., Refregier, A., \& Ellis, R. 2002, \mnras, 344, 673 (astro-ph/0203134)

\bibitem[Bardeen, Bond, Kaiser, \& Szalay(1986)]{BBKS}Bardeen, J.~M., Bond, J.~R., Kaiser, N., \& Szalay, A.~S.\ 1986, \apj, 304, 15

\bibitem[Bennett et al.(1996)]{Bennett96} Bennett, C.~L.~et al.\ 1996, \apjl, 464, L1

\bibitem[Birkinshaw, Gull \& Hardebeck(1984)]{Birkinshaw84} 
Birkinshaw, M., Gull, S.~F., Hardebeck, H. 1984, \nat, 309, 34

\bibitem[Bond(1988)]{Bond88} Bond, J.~R. 1988, in {\it The Early
Universe}, p. 283-334, ed. W. Unruh and G.  Semenoff, Dordrecht:Reidel.

\bibitem[Bond(1994)]{yukawa93}Bond, J.~R. 1994, in { Relativistic
Cosmology}, ed., M. Sasaki, Proc. 8th Nishinomiya-Yukawa Memorial
Symposium (Universal Academy Press), pp. 23-55, (astro-ph/9406075)

\bibitem[Bond(1996)]{bh95}Bond, J.~R. 1996,  in {Cosmology and
Large Scale Structure}, Les Houches Session LX, (ed. R. Schaeffer, J. Silk, M.
Spiro \& J. Zinn-Justin),  pp.~469-674. Elsevier. (Available at   
http://www.cita.utoronto.ca/$\sim$bond/papers/houches/)

\bibitem[Bond \& Crittenden(2002)]{Bond02}Bond, J.~R. \& Crittenden,
R.~G., 2002, in {Structure Formation in the Universe}, 241-280,
R.G. Crittenden and N.G. Turok, eds., Proc. NATO ASI (Kluwer Academic
Publishers) (astro-ph/0108204)

\bibitem[Bond \& Efstathiou(1987)]{Bond87}Bond, J.~R.~\& Efstathiou, G.\ 1987, \mnras, 226, 655

\bibitem[Bond \& Jaffe(1999)]{Bond99}Bond, J.~R.~\& Jaffe, A.~H.\
1999, {Large-Scale Structure in the Universe}, Phil. Trans. R. Soc. Lond. A 357, 57 ( astro-ph/9809043)

\bibitem[Bond, Jaffe \& Knox(2000)]{BJK2000}
Bond, J. R., Jaffe, A. H., \& Knox, L. 2000, \apj, 533, 19

\bibitem[Bond et al.(1998)]{Bond98}Bond, J.~R., Kofman, L., Pogosyan, D. \& Wadsley, J. 1998, in {Wide Field Surveys in Cosmology},
Proc. XIV IAP Colloquium, p. 17, (ed, Colombi, S. \& Mellier, Y.), Paris:
Editions Frontieres (astro-ph/9810093)

\bibitem[Bond \& Myers(1996)]{BM96}Bond, J.~R. \& Myers, S. 1996, \apjs, 103, 1

\bibitem[Bond et al.(2002)]{Bond01}Bond, J.~R., Ruetalo, M.~I.,
Wadsley, J.~W. \& Gladders, M.~D., 2002, Astron. Soc. Pacific Conference
Series 257, 327-339, {AMiBA 2001: High-z Clusters,
Missing Baryons, and CMB Polarization}, Proc. TAW8, Taiwan, June,
ed. L-W Chen, C-P Ma, K-W Ng and U-L Pen, (San Francisco:
Astron. Soc. Pacific) (astro-ph/0112499)

\bibitem[Bond et al.(2003)]{broysoc03}Bond, J.~R., Contaldi, C.~R. \& Pogosyan, D. 2003, Phil. Trans. R. Soc. Lond. A 361, 2435  (astro-ph/0310735)

\bibitem[Borgani et al.(2001)]{Borgani01}Borgani, S., et al. 2001, \apj, 561, 13 (astro-ph/0106428) 

\bibitem[Brown et al.(2003)]{Brown03}Brown, M., Taylor A., Bacon D.,
Gray M., Dye S., Meisenheimer K. \& Wolf C. 2003, \mnras, 341, 100
(astro-ph/0210213)

\bibitem[Carlberg et al.(1997)]{Carlberg97}Carlberg, R.~G., Morris,
S.~L., Yee, H.~K.~C., Ellingson, E. 1997, \apj, 479, L19

\bibitem[Carlstrom et al.(1996)]{Carlstrom96} Carlstrom, J.~E. et al. 1996, \apj, 456, L75

\bibitem[Cole \& Kaiser(1988)]{Cole88}Cole, S. \& Kaiser, N., 1988,
    \mnras, 233, 637

\bibitem[Cooray(2000)]{Cooray00}Cooray, A., 2000, \prd, 62, 103506

\bibitem[Cooray \& Sheth(2002)]{CooraySheth02}Cooray, A. \& Sheth, R. 2002, 
Physics Reports 372, 1

\bibitem[Dawson et al.(2002)]{BIMA02}Dawson, K.S., Holzapfel, W.L., Carlstrom, J.E.,
LaRoque, S.J., Miller, A., Nagai, D. \& Joy, M. 2002, \apj\ 581, 86 (astro-ph/0206012)

\bibitem[Dodelson et al.(2001)]{Dodelson01}Dodelson, S., et al.~2001, \apj, 572, 140  (astro-ph/0107421)

\bibitem[Efstathiou et al.(1992)]{ebw}Efstathiou, G., Bond, J.~R.  \& White, S.~D.~M. 1992,  \mnras, 258, 1P

\bibitem[Eke et al.(1996)]{Eke96}Eke, V.~R., Cole, S., Frenk, C.~S. 1996, \mnras, 282, 263

\bibitem[Fan \& Bahcall(1998)]{Fan98}Fan, X.  \& Bahcall, N.~A. 1998, \apj, 504, 1

\bibitem[Goldstein et al.(2003)]{Goldstein02}Goldstein J., et al. 2003, \apj, 599, 773 (astro-ph/0212517)

\bibitem[Grainge et al. (2002)]{Grainge02}Grainge, K., et al. 2002, \mnras, 329, 890

\bibitem[Grainge et al. (2003)]{VSAext02}Grainge, K., et al. 2003,
  \mnras, 341, L23

\bibitem[Halverson et al.(2001)]{DASI}Halverson, N.~W.~et al.\ 2002, \apj, 568, 38

\bibitem[Hamana et al.(2003)]{Hamana03}Hamana, T. et al.  2003, \apj,
  597, 98 (astro-ph/0210450)

\bibitem[Heymans et al.(2004)]{Heymans03}Heymans, C., Brown, M. Heavens, A., Meisenheimer, K. \& Taylor, A. \&
 Wolf, C.  2004, \mnras, 347, 895 (astro-ph/0310174)

\bibitem[Holzapfel et al.(1997)]{Holzapfel97}Holzapfel, W.~L. et al. 1997, \apj, 479, 17

\bibitem[Hoekstra et al.(2002)]{Hoekstra02}Hoekstra, H., Yee, H.~K.~C.
\& Gladders, M.~D. 2002, \apj,  577, 595 (astro-ph/0204295)

\bibitem[Jaffe et al.(2001)]{Jaffe01}Jaffe, A.~H. et al.~2001, \prl, 86, 3475

\bibitem[Jarvis et al.(2003)]{Jarvis03}Jarvis, M., Bernstein, G., Fischer, P., Smith, D., Jain, B., Tyson, A. \& Wittman, D., 2003, AJ, 1025, 1014

\bibitem[Kogut et al.(2003)]{Kogut03}Kogut, A. et al., 2003, \apjs,  148, 161 

\bibitem[Komatsu \& Kitayama(1999)]{Komatsu99}Komatsu, E. \& Kitayama, T., 1999, \apjl, 526, L1

\bibitem[Komatsu \& Seljak(2002)]{Komatsu:2002wc}Komatsu, E. \& Seljak, U. 2002, \mnras, 336, 1256 
(astro-ph/0205468)

\bibitem[Kuo et al.(2004)]{Kuo02}Kuo, C. L. et al., 2004, \apj, 600, 32 (astro-ph/0212289)

\bibitem[Lahav et al.(2002)]{Lahav02}Lahav, O., et al.~2002, \mnras, 333, 961 (astro-ph/0112162)

\bibitem[Lange et al.(2001)]{Lange01}Lange, A.~E., et al.~2001, \prd, 63, 042001

\bibitem[Lee et al.(2001)]{Lee01}Lee, A.~T., et al. 2001, \apj, 561, L1

\bibitem[Makino \& Suto(1993)]{Makino93}Makino, N., \& Suto, Y. 1993, \apj, 405, 1

\bibitem[Mason,  Myers, \& Readhead(2001)]{Mason01}Mason, B.~S., Myers, S.~T., \& Readhead, A.~C.~S. 2001, \apj, 555, L11

\bibitem[Mason et al.(2003)]{Mason02}Mason, B.~S., et al.~2003, \apj, 591, 540 (Paper II)

\bibitem[Mauskopf et al.(2000)]{B97} Mauskopf, P.~D., et al.~2000, \apj, 536, L59

\bibitem[Melchiorri \& Silk(2002)]{Melchiorri02}Melchiorri, A. \& Silk, J., 2002 Phys. Rev. D 66, 041301 (astro-ph/0203200)

\bibitem[Miller et al.(1999)]{TOCO} Miller, A.~D., et al.~1999, \apjl, 524, L1

\bibitem[Mo \& White(1996)]{Mo96}Mo, H.~J. \& White, D.~M., 1996, \mnras, 282, 347

\bibitem[Molnar \& Birkinshaw(2000)]{Molnar00}Molnar, S. \& Birkinshaw, 2000, \apj, 537, 542

\bibitem[Myers et al.(2003)]{Myers02}Myers, S.~T. et al.~2003, \apj, 591, 575 (Paper IV)

\bibitem[Netterfield et al.(2002)]{Netterfield02}Netterfield, C.~B., et al.~2002, \apj\ 571, 604 (astro-ph/0104460)

\bibitem[Padin et al.(2001)]{Padin01}Padin, S., et al.~2001, \apjl, 549, L1 (Paper I)

\bibitem[Peacock \& Dodds(1994)]{Peacock94}Peacock, J.~A.~\& Dodds, S.~J.\ 1994, \mnras, 267, 1020

\bibitem[Peacock et al.(2001)]{Peacock01}Peacock, J.~A., et al.\ 2001, \nat, 410, 169

\bibitem[Pearson et al.(2003)]{Pearson02}Pearson, T.~J., et al.~2003, \apj, 591, 556 (Paper III)

\bibitem[Peebles \& Yu(1970)]{Peebles70}Peebles, P.~J.~E.~\& Yu, J.~T.\ 1970, \apj, 162, 815

\bibitem[Pen(1998a)]{Pen98a}Pen, U.-L., 1998a, \apjs, 115, 19

\bibitem[Pen(1998b)]{Pen98b}Pen, U.-L., 1998b, \apj, 498, 60

\bibitem[Pen(1998c)]{Pen98c}Pen, U.-L. 1998c, \apj, 504, 60

\bibitem[Pierpaoli, Scott \& White(2001)]{Pierpaoli01}Pierpaoli, E., Scott, D., \& White, M. 2001, \mnras, 325, 77

\bibitem[Pierpaoli et al.(2003)]{Pierpaoli03}Pierpaoli, E., Borgani, S., Scott, D., \& White, M. 2003, \mnras, 342, 163
(astro-ph/0210567)
 
\bibitem[Press \& Schechter(1974)]{Press74}Press, W.~H.~\& 
Schechter, P.\ 1974, \apj, 187, 425

\bibitem[Pryke et al.(2002)]{Pryke01}Pryke, C., Halverson, 
N.~W., Leitch, E.~M., Kovac, J., Carlstrom, J.~E., Holzapfel, W.~L., \& 
Dragovan, M.\ 2002, \apj, 568, 46

\bibitem[Readhead et al.(2004)]{Readhead04}Readhead, A.~C.~R., et al.~2004, \apj, 609, 498

\bibitem[Refregier et al.(2002)]{Refregier02}Refregier, A., Rhodes, J., \& Groth, E.~J. 2002, \apjl, 572, L131 (astro-ph/0203131)

\bibitem[Reiprich \& B{\"o}hringer(2002)]{Reiprich01}Reiprich, T.~H. \& B{\"o}hringer, H. 2002, \apj, 567, 716 (astro-ph/0111285)

\bibitem[Ruhl et al.(2003)]{Ruhl02}Ruhl, J.E. et al., 2003, \apj, 599, 786 (astro-ph/0212229)

\bibitem[Schuecker et al.(2003)]{SBCG03}Schuecker P., Bohringer H., Collins C.A \&  Guzzo L. 2003, A\&A, 398, 867 (astro-ph/0208251)

\bibitem[Scott et al.(2002)]{VSA02}Scott, P.F. et al. 2002, \mnras\ 341, 1076 (astro-ph/0205380)

\bibitem[Seljak(2002)]{Seljak01}Seljak, U. 2002, \mnras, 337, 769 (astro-ph/0111362)

\bibitem[Seljak, Burwell \& Pen(2001)]{Seljak01b} Seljak, U., Burwell, J. \& Pen, U.-L., 2001, \prd, 63, 063001

\bibitem[Sievers et al.(2003)]{Sievers02}Sievers, J.~L., et al.~2003, \apj, 591, 599 (Paper V)

\bibitem[Silk(1968)]{Silk68}Silk, J.\ 1968, \apj, 151, 459

\bibitem[Spergel et al.(2003)]{Spergel03}Spergel, D.~N., et al., 2003, \apjs,  148, 175 

\bibitem[Springel, White \& Hernquist(2001)]{Springel01}Springel, V., White, M., \& Hernquist, L.\ 2001, \apj, 549, 681,
erratum; Springel, V., White, M., \& Hernquist, L.\ 2001, \apj, 562, 1086

\bibitem[Sugiyama(1995)]{sugiyama94supp}Sugiyama, N. 1995, \apjs. 100, 281

\bibitem[Sunyaev \& Zeldovich(1970)]{SUNYAEV}Sunyaev, R.~A. \& Zeldovich, Ya.~B. 1970, \apss, 7, 3

\bibitem[Szalay et al.(2001)]{Szalay01}Szalay, A.~S. et al. 2001 , \apj,  591, 1 (astro-ph/0107419)

\bibitem[Udomprasert et al.(2004)]{Udomprasert01}Udomprasert, P.~S.,
Mason, B.~S., Readhead, A.~C.~S., \& Pearson, T. J.  2004, \apj, 615, 63

\bibitem[Verde et al.(2002)]{Verde01}Verde, L. et al. 2002, \mnras,
  335, 432 (astro-ph/0112161)

\bibitem[Viana, Nichol \& Liddle(2001)]{Viana01}Viana, P.~T~.P., Nichol,
R.~C. \& Liddle, A.~R. 2001, \apjl,  569, L75 (astro-ph/0111394)

\bibitem[Voevodkin \& Vikhlinin(2004)]{VV03}Voevodkin, A. \& Vikhlinin, A. 2004, \apj, 601, 610 (astro-ph/030554) 

\bibitem[Wadsley, Stadel \& Quinn(2003)]{Wadsley03}Wadsley, J.~W., Stadel, J. \& Quinn, T. 2003,  New Astronomy, 9, 137

\bibitem[van Waerbeke et al.(2002)]{Ludo02}van Waerbeke, L., Mellier,
Y., Pello, R., Pen, U.-L, McCracken, H.~J., \& Jain, B.  2002, A\&A, 393, 369 (astro-ph/0202503)

\bibitem[van Waerbeke et al.(2005)]{Ludo05}van Waerbeke, L., Mellier,
Y., \& Hoekstra, H.  2005, A\&A, 429, 75


\bibitem[Zhang \& Pen(2001)]{Zhang01}Zhang, P.~J., \& Pen, U.-L., 2001, \apj, 549, 18

\bibitem[Zhang, Pen \& Wang(2002)]{Zhang02}Zhang, P.~J., Pen, U.-L. \&
    Wang, B. 2002, \apj, 577, 555  (astro-ph/0201375)

\end{thebibliography}
\end{document}